\documentclass[acmtog]{acmart}

\copyrightyear{2026}
\acmYear{2026}
\setcopyright{cc}
\setcctype{by}
\acmJournal{TOG}
\acmBooktitle{Craig-Boris '25: Craig's test event for Boris LaTeX Code 2 (Craig-Boris '25), January 1--5, 2025, TBa, TAS, USA}
\acmDOI{10.1145/3563045.3563055}
\acmISBN{/25/01}

\usepackage{graphicx}
\usepackage{subfig}
\usepackage{xcolor}
\usepackage{multirow}
\usepackage{float}
\usepackage{makecell}

\usepackage[ruled]{algorithm2e} 

\SetAlFnt{\small}
\SetAlCapFnt{\small}
\SetAlCapNameFnt{\small}
\SetAlCapHSkip{0pt}

\usepackage{enumitem}

\newcommand{\head}[1]{\vspace{2mm}\noindent\textbf{#1}}
\newcommand{\method}{{TG-Diff}}

\begin{document}


\begin{teaserfigure}
\centering
  \includegraphics[width=\linewidth]{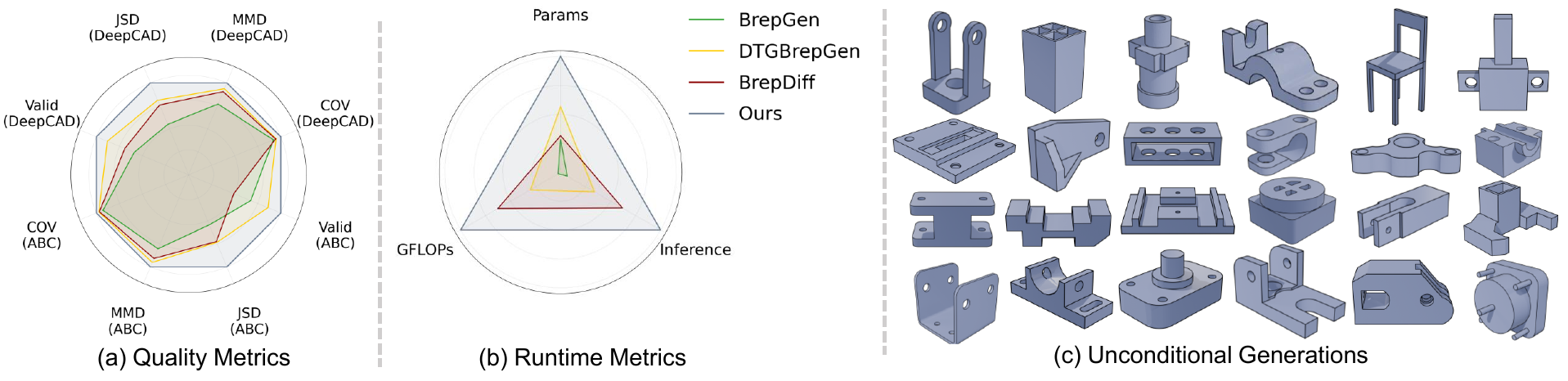}
  \caption{Overview of B-rep generation results. (a) The proposed \method{} consistently outperforms the competing methods across all geometric quality metrics, including validity (valid), COV, MMD, and JSD on both the DeepCAD-0-30 dataset and the ABC-0-50 dataset. (b)
  \method{} demonstrates superior runtime efficiency (time, GFLOPs, Params). (c) Our framework can produce structurally diverse B-rep models in unconditional generation scenarios.}
  \label{fig:teaser}
\end{teaserfigure}



\author{MingZe Sun}
\orcid{0009-0004-0336-6339}
\email{1834521663@qq.com}
\affiliation{%
  \institution{University of Chinese Academy of Sciences}
  \city{Beijing}
  \country{China}
}

\author{Haiyong Jiang}
\orcid{0000-0001-7348-5844}
\email{haiyong.jiang@ucas.ac.cn}
\affiliation{%
  \institution{University of Chinese Academy of Sciences}
  \department{School of Artificial Intelligence}
  \city{Beijing}
  \country{China}
}

\author{Bingchen Yang}
\email{bingchenyang@ntu.edu.sg}
\affiliation{%
  \institution{Nanyang Technological University}
  \city{Singapore}
  \country{Singapore}
}

\author{Haoxuan Song}
\email{songhaoxuan241@mails.ucas.ac.cn}
\affiliation{%
  \institution{University of Chinese Academy of Sciences}
  \city{Beijing}
  \country{China}
}

\author{Yidi Li}
\email{liyidi@ucas.ac.cn}
\affiliation{%
  \institution{University of Chinese Academy of Sciences}
  \city{Beijing}
  \country{China}
}

\author{Jun Xiao}
\email{xiaojun@ucas.ac.cn}
\affiliation{%
  \institution{University of Chinese Academy of Sciences}
  \city{Beijing}
  \country{China}
}

\author{Peter Wonka}
\email{pwonka@gmail.com}
\affiliation{%
  \institution{King Abdullah University of Science and Technology (KAUST)}
  \country{Saudi Arabia}
}



\title{\method{}: Coupling Discrete Topology Diffusion and Topology-conditioned Geometry Diffusions for B-Rep Generation}





\begin{abstract}
Boundary representation (B-rep) is the standard format for computer-aided design (CAD).
This article proposes a lightweight two-stage diffusion-based B-rep generation framework \method{} that achieves efficient, high-quality B-rep generation by decoupling topology and geometric modeling.
In contrast to previous work that generates topology as a collection of vertices, edges, and surfaces and their relationships, \method{} represents the topology only as a collection of surfaces and their adjacency relationships.
This surface-centric representation inherently alleviates the geometric and topological inconsistencies between separately generated surfaces, edges, and vertices, simplifying the generation process.
Based on the surface-centric representation, we develop two independent diffusion models that generate surface adjacency relationships and surface latents, respectively. By using topology as guidance, the surface generation process becomes more stable, leading to stronger structural completeness in the generated B-rep models.
The topology diffusion model adopts a Discrete Diffusion Model (D3PM) for efficient binary sampling, avoiding the slow inference of autoregressive methods. 
The surface latent generation adapts a conditional latent diffusion model with a lightweight DiT architecture, where surface adjacency guides geometry generation while reducing computational cost.
Finally, edges and vertices are derived from the decoded adjacent surfaces via post-processing to form a final watertight B-rep.
Despite its compact computational footprint (82.18M parameters and 2.2GFLOPs), \method{} excels in the validity metric while achieving superior performance on all COV, MMD, and JSD  metrics across the DeepCAD and the ABC dataset.
\end{abstract}


\maketitle

\begin{table*}[t]
\centering
\caption{Comparison of B-rep generation pipelines.}
\label{tab:comparison_pipeline}
\small
\setlength{\tabcolsep}{4pt}
\begin{tabular}{l|cc|ccc}
\toprule
& \multicolumn{2}{c|}{Topology Generation} 
& \multicolumn{3}{c}{Geometry Generation} \\
\cmidrule(lr){2-3} \cmidrule(lr){4-6}
& Topology Representation & Architecture & Geometry Representation  & Diffusion Stages & W/ Topo Cond \\\midrule
BrepGen & N/A & N/A  & Surface, edge, vertex  & 4 & $\times$ \\
DTGBrepGen & Surfaces, edge, vertex adjacency & AutoRegressive  & Surface, edge, vertex & 4 & $\checkmark$ \\
BrepDiff & N/A & N/A & Surface & 1 & $\times$ \\
Hola & N/A & N/A  & Surface  & 1 & $\times$ \\
Ours & Surfaces adjacency & D3PM & Surface & 1 & $\checkmark$ \\
\bottomrule
\end{tabular}
\end{table*}

\section{Introduction} \label{sec:intro}
Boundary representation (B-rep) is the de facto industry standard for 3D Computer-Aided Design (CAD), valued for its compactness, analytic geometric precision, and topological flexibility during editing. 
Despite these advantages, B-rep generation remains a formidable challenge. 

Initial exploration on B-rep generation favored autoregressive (AR) models.
AR-based methods~\cite{para21sketchgen,xu22skexgen,jaya23solidgen,li25brepgpt, xu25autobrep}, while being compatible with LLMs, often suffer from quantization limitations and accumulated sequence errors.

More recently, diffusion-based models are being used. We can distinguish two sets of diffusion approaches.
The first set of methods generates the geometry first and recovers the topology in a post-process. In this context, BrepGen~\cite{xu24brepgen} pioneered the explicit generation of a geometric hierarchy comprising vertices, curves, and surfaces via a multi-stage diffusion process. HoLa~\cite{liu25hola} learns a latent distribution via a variational autoencoder (VAE), enabling the joint generation of surfaces and curves while implicitly capturing topological relationships.
Similarly, BrepDiff~\cite{lee25brepdiff} adopts a single-stage diffusion model to directly generate surfaces, with topology inferred afterward through mesh intersection–based post-processing, but without explicitly enforcing topological consistency during generation. The main disadvantage of these methods is that B-Rep generation is a heavily constrained problem, and it is easy to have failed generations due to inconsistencies and misalignments between surfaces, edges, and vertices.
To combat this difficulty, the second set of methods generates the topology first, and uses the topology as a condition for geometry generation. 
DTGBrepGen~\cite{li25dtg} introduces a topology-first strategy by explicitly generating surface–edge and edge–vertex relationships using an autoregressive (AR) scheme prior to geometry synthesis, thereby improving structural consistency. However, explicitly generating vertices, edges, and surfaces leads to two key limitations: 1) The topology description is overly complex and redundant, and the topology itself can become inconsistent. 2) The redundant description in topology and geometry leads to bigger-than-necessary latent spaces in both generation phases. Our work directly addresses these two limitations.

To this end, we introduce \method{}, a two-stage diffusion framework for B-rep generation. 
Compared to DTGBrepGen, we propose to eliminate vertices and edges from the topology (and geometry) representation during diffusion and employ a surface-only B-rep topology and geometry representation.
The topology diffusion learns the distribution of surface adjacency as topology using a discrete diffusion model. 
Subsequently, we encode B-rep surfaces into a compact latent space using a topology-aware variational autoencoder.
Afterward, we extend the Diffusion Transformer (DiT)~\cite{ref/DiT} with a topology modulation and lightweight representation learning using a six-layer low-dimensional DiT layer and a two-layer high-dimensional DiT layer. 
Finally, the predicted topology and B-rep surfaces can be postprocessed to form an integrated B-rep shape with surface trimming and stitching. 
In our approach, vertices and edges are not explicitly generated but are induced via surface intersections during post-processing.

Adopting a surface-centric approach offers several critical advantages.
First, focusing on surfaces significantly reduces the diffusion space complexity, as the number of surfaces is significantly smaller than the number of edges and vertices. A compact latent space is a major contributor to a successful diffusion model.
Second, a surface-centric representation inherently mitigates geometric-topological misalignment. An independently generated edge might not lie precisely on its corresponding surface. By deriving edges directly from surface intersections, we ensure that every edge is, by definition, perfectly located within its parent surfaces.
Furthermore, modeling the full B-rep hierarchy requires maintaining consistency across multiple levels (Face-Loop-Edge-Vertex). A surface-centric approach simplifies this into a single adjacency graph, preventing orphan geometric entities.
We show the conceptual comparisons of these methods in Tab.~\ref{tab:comparison_pipeline}.

Experimental results demonstrate the superiority of our method over previous works.
Moreover, the runtime statistics show that the method is much more efficient in both parameters and running time. See Fig.~\ref{fig:teaser} for an overview of generation results.

\begin{figure*}[t]
    \centering
    \includegraphics[width=\linewidth]{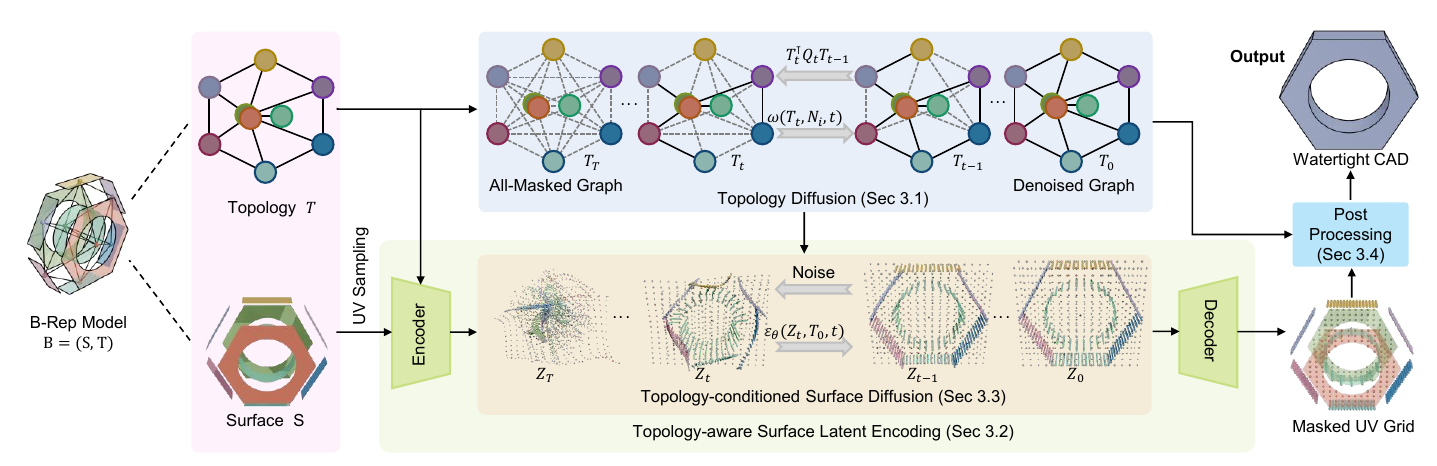}
    \caption{The pipeline of the method. We represent a B-rep model as masked UV grid surfaces $\mathbf{S}$ and their mutual adjacency as a topology matrix $\mathbf{T}$. The B-rep generation consists of a topology diffusion process for topology generation, a topology-aware surface latent encoding network, and a topology-conditioned surface diffusion for surface generation. The generated topology and surfaces are post-processed to form watertight B-rep models. Note that the encoded surface latents are denoted as $\mathbf{Z}$.}
    \label{fig:pipeline}
\end{figure*}


\section{Related Works} \label{Sec: Related Works}
Boundary representation and CAD command sequences stand as two primary formats for structural CAD shapes.
This section reviews generative models for B-rep models and CAD command sequences.

\subsection{B-rep Generation}\label{subsec:b-rep-works}
B-rep models comprise a hierarchical assembly of geometric entities, including vertices, edges, and surfaces, coupled with complex topological relationships that define their connectivity. 
This intricate interdependence between geometry and topology renders B-Rep generation significantly formidable. 
Depending on the underlying generative models, we categorize existing research into autoregressive and diffusion-based B-rep generation.

Autoregressive-based B-rep generation frames the task as a sequential generation of B-rep geometric entities and topologies. 
As a pioneering effort, SolidGen~\cite{jaya23solidgen} progressively constructs B-Reps by predicting vertices, edges, and surfaces using separate networks. 
SolidGen encodes topological connectivity by generating each hierarchical level (e.g., surfaces) contingent upon the preceding geometric entities (e.g., edges and vertices) via a pointer network mechanism~\cite{vinyals15pointer}.
Stitch-a-Shape~\cite{li25stichshape} extends SolidGen by interleaving explicit topology-prediction modules within the geometric generation process, thereby constraining subsequent geometric stages with previously inferred topology.
AutoBrep~\cite{xu25autobrep} presents a novel topology tokenization that encodes surface-edge connections as a local reference to surfaces, allowing for a unified sequence of discrete tokens for autoregressive predictions. 
BrepGPT~\cite{li25brepgpt} decomposes B-rep models into unified local Voronoi Half-Patches (VHPs) and then generates VHPs autoregressively.
CMT~\cite{wu2025cmt} generates edge and surface latents using a MAR model~\cite{li24mar}, respectively, and employs cross-attention over these latent features to predict the surface–edge adjacency.
BrepARG~\cite{li2026breparg} represents B-rep geometry and topology as a unified token sequence and applies a decoder-only Transformer for autoregressive B-rep generation.
Although AR methods achieve strong performance, their discrete token representation and sequential generation limit geometric precision and inference efficiency.

Another category of works is based on diffusion models~\cite{ho20ddpm, song21ddim}. 
The seminal work BrepGen~\cite{xu24brepgen} presents a top-down generation strategy, first generating surfaces and then sequentially predicting the boundary edges and vertices for each surface. A post-processing step subsequently merges duplicated edges and vertices to stitch the generated primitives into a complete B-rep.
DTGBrepGen~\cite{li25dtg} augments BrepGen by conditioning the multi-stage generation process on an autoregressively generated surface adjacency graph.
HoLa~\cite{liu25hola} represents B-rep primitives and their topological dependencies through a unified set of holistic latent codes, enabling effective diffusion-based generation. 
BrepGiff~\cite{guo25brepgiff} employs an edge-guided discrete graph diffusion model~\cite{chen23edge} to generate each surface and its connection with other surfaces progressively.
BrepDiff~\cite{lee25brepdiff} employs a single-stage diffusion model with a masked UV-grid surface representation to mitigate non-deterministic dependencies between multi-stage generation.
BR-DF~\cite{zhang2025brdf} presents volumetric distance functions for B-reps, avoiding the generation of complex topology structures.
GraphBRep~\cite{li2025graphbrep} adopts a multi-stage generation framework for B-Rep modeling. It first generates surface geometries, then predicts surface adjacency relationships through graph diffusion, and finally synthesizes edge geometries based on the generated topology graph, improving topology generation efficiency and reducing redundant edge representations.
This work presents a two-stage diffusion framework for B-rep surface and topology generation. 

\subsection{CAD Command Sequence Generation}\label{subsec:cadseq}
Learning the generation of the CAD command process goes beyond B-rep shapes and enables the understanding of inherent design intents. 
DeepCAD~\cite{wu21deepcad} presents a seminal CAD dataset and organizes CAD operations as a common structure for generative CAD learning.
Following works improve generative models by introducing geometry-aware tokenization and autoregressive models~\cite{xu22skexgen, xu23hncad}.
Different from SkexGen, which autoregressively generates sketch-and-extrude CAD construction sequences, SketchDNN~\cite{chereddy2025sketchdnn} uses diffusion to generate 2D sketch primitives with both discrete types and continuous geometry.
Thereafter, the emergence of Large Language Models (LLMs) motivates recent explorations of LLM-based CAD sequence generation~\cite{xu24cadmllm, wang25cadgpt, zhang25flexcad, li25cadllama, you24img2cad}.
However, annotated CAD command datasets suitable for model training or fine-tuning remain scarce, and most existing datasets cover only a limited range of command types, primarily sketch and extrude commands.
This limitation substantially restricts the diversity and expressiveness of the generated CAD models.

\subsection{Primitive Fitting for B-rep Generation}
In contrast to native B-rep generation, another choice is to construct B-rep shapes from 3D shapes generated by off-the-shelf 3D generative models~\cite{long24wonder3d,xiang25trellis}.
This eases the learning of 3D generation, but the additional primitive fitting and connection are still quite challenging and prone to errors.
Previous methods explore different kinds of primitive fitting strategies, including detection-based methods~\cite{guo22complexgen, li25caddreamer}, implicit surface fitting for clustered points~\cite{liu24point2cad}, Voronoi-based geometric decomposition~\cite{liu24splitandfit}, or direct primitive prediction from point clouds~\cite{ma25point2primitive}.
While conceptually straightforward, such a pipeline is sensitive to geometric noise and involves time-consuming optimization for fine-grained topological consistency.

\section{Methodology} \label{sec:method}
We aim at the unconditional generation of B-rep models using a surface-centric representation. 
This representation prioritizes surface topology and geometry, avoiding the complexity of explicitly modeling multi-level vertex and edge hierarchies, and directly uses the generated surfaces to intersect and obtain edges and vertices via post-processing.

Formally, we represent a B-rep model as $\mathbf{B} = (\mathbf{S}, \mathbf{T})$ with $\mathbf{S}$ denoting the set of $N$ parametric surfaces (geometry) and $\mathbf{T} \in \{0,1,2\}^{N \times N}$ denoting the surface adjacency relationships (topology). 
$\mathbf{T}_{ij}$ represents a categorical variable with $K=3$ states. 
Specifically, $\mathbf{T}_{ij} = 1$ indicates that surface $i$ and surface $j$ are adjacent (i.e., they intersect at a shared edge), $\mathbf{T}_{ij} = 0$ denotes no such adjacency, and $\mathbf{T}_{ij} = 2$ represents the padding token. During batch training, the model is also required to identify and predict padding positions.
Following BrepGen~\cite{xu24brepgen}, each surface $\mathbf{S} \in \mathbb{R}^{N\times 16 \times 16 \times 4}$ is represented with a uniformly sampled masked UV grid of resolution $16 \times 16$.
The four channels include 3D coordinates $\mathbf{S}_\text{coord}$ for each grid point and a binary mask $\mathbf{S}_\text{mask}$ indicating whether a grid point lies on the final trimmed B-rep surface.

Based on the surface-centric representation, we decouple B-rep generation into discrete topology generation and surface geometry generation.
As illustrated in Fig.~\ref{fig:pipeline}, the framework first learns the generation of surface topology $\mathbf{T}$ via a discrete diffusion process. 
Subsequently, we learn the encoding of surface latents and generate surface latents through a topology-conditioned latent diffusion model. 
The diffusion process is conditioned on the generated topology $\mathbf{T}$ via self-attention-based modulation to enhance global structural awareness.
Finally, parametric surfaces decoded from generated surface latents are post-processed to form watertight B-reps.

\subsection{Topology Diffusion}\label{subsec:topo-diff}
We use Discrete Diffusion Probabilistic Models (D3PM)~\cite{austin21d3pm} for topology generation, to accommodate the discrete nature of adjacency relationships.
The diffusion process consists of a forward and a reverse diffusion process. 

The forward diffusion process at timestep $t$ is defined as:
$q(\mathbf{T}_t \mid \mathbf{T}_{t-1}) := \mathbf{T}_{t}^\top \mathbf{Q}_t \mathbf{T}_{t-1}$,
where $\mathbf{Q}_t$ denotes the transition probability matrix between topology states. 
Then the marginal probability can be derived as:
\begin{equation}
    q(\mathbf{T}_t \mid \mathbf{T}_0)
    := \mathbf{T}_t^\top \bar{\mathbf{Q}}_t \mathbf{T}_0, 
    \qquad \bar{\mathbf{Q}}_t = \mathbf{Q}_t \mathbf{Q}_{t-1} \cdots \mathbf{Q}_1. 
    \label{eq:topo-forward-diff-marginal}
\end{equation}
Meanwhile, the backward process of the framework can be described as:
\begin{equation}
    q(\mathbf{T}_{t-1} \mid \mathbf{T}_t, \mathbf{T}_0)
    :=
    \frac{
        \big(
            \mathbf{T}_t^\top \mathbf{Q}_t \mathbf{T}_{t-1}
        \big)
        \big(
            \mathbf{T}_{t-1}^\top \bar{\mathbf{Q}}_{t-1} \mathbf{T}_0
        \big)
    }{
        \mathbf{T}_t^\top \bar{\mathbf{Q}}_t \mathbf{T}_0
    },
    \label{eq:topo-posterior}
\end{equation}
Following VQ-Diffusion~\cite{gu2022vector}, we introduce an additional absorbing \texttt{[MASK]} state into the transition matrix to explicitly indicate corrupted positions, resulting in $K+1$ discrete states.

$\mathbf{Q}_t \in \mathbb{R}^{(K+1)\times(K+1)}$ can be formulated as:
\begin{equation}
\mathbf{Q}_t =
\begin{bmatrix}
\alpha_t \mathbf{I}_K + \beta_t \mathbf{1}_K \mathbf{1}_K^\top & \mathbf{0} \\
\gamma_t \mathbf{1}_K^\top & 1
\end{bmatrix},
\label{eq:vqdiffusion-transition}
\end{equation}
where $\mathbf{I}_K \in \mathbb{R}^{K \times K}$ denotes the identity matrix and
$\mathbf{1}_K \in \mathbb{R}^{K}$ is the all-ones vector. 

For $\mathbf{Q}_t \in \mathbb{R}^{(K+1)\times(K+1)}$, the transition probability of replacing a non-mask state with the \texttt{[MASK]} state is $\gamma_t$.
The transition probability between different categorical states is $\beta_t$.
Consequently, an entity remains unchanged with probability $\alpha_t = 1 - K\beta_t - \gamma_t$.
The \texttt{[MASK]} itself always preserves its state. 

The reverse diffusion step $p_{\theta_1}(\mathbf{T}_{t-1}|\mathbf{T}_t)$ maps a noisy topology $\mathbf{T}_{t}$ to a less noisy topology $\mathbf{T}_{t-1}$ via a denoising network $\omega_{\theta_1}(\mathbf{T}_{t}, N_i, t)$ that is conditioned on the surface count $N_i$ and the time step $t$.
To accelerate and stabilize training, we use BLISS~\cite{bliss07,bliss11} for canonical labeling of the surface sequences, which reduces permutation ambiguity in advance.

\begin{figure}[h]
    \centering
    \includegraphics[width=1.0\linewidth]{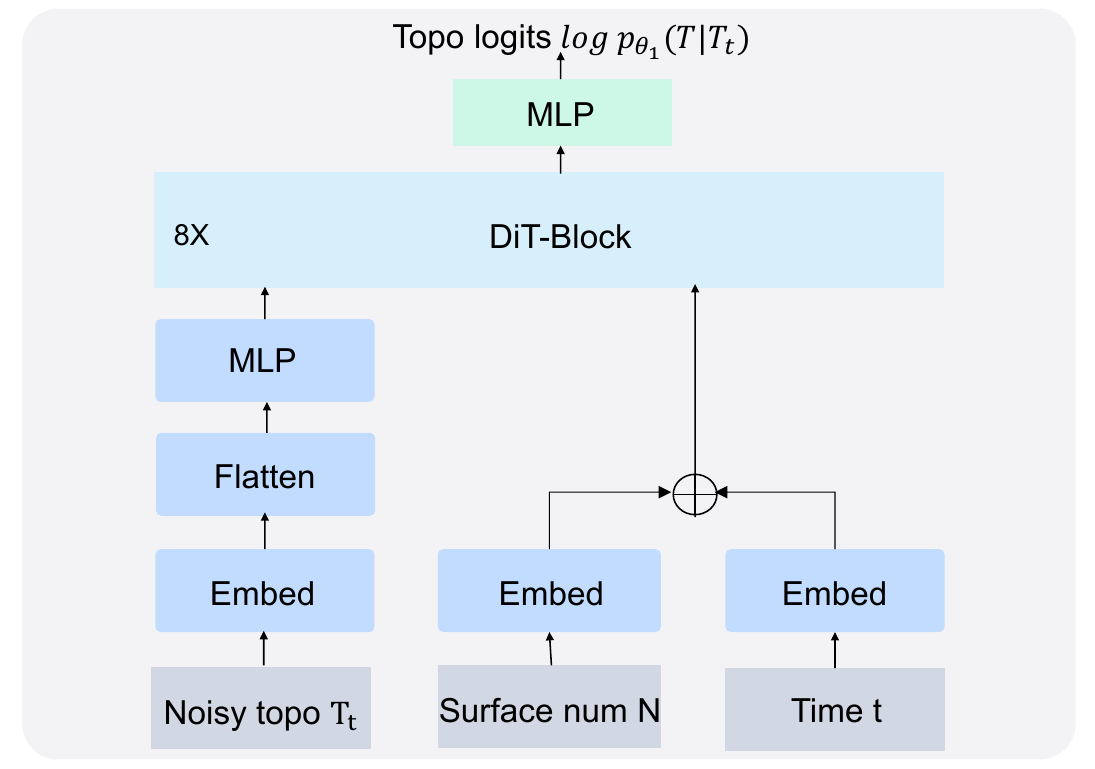}
    \caption{The architecture for the topology denoising network. Flatten performs a flattening operation on the last two dimensions of a tensor.}
    \label{fig:topo_diffusion}
\end{figure}

The topological denoising network is illustrated in Fig.~\ref{fig:topo_diffusion}.
It begins by mapping $\mathbf{T}_t$ to an initial embedding with De=$8$ channels. 
Here, $\mathbf{T}_t$ denotes the noisy topology at diffusion step $t$.
Rather than treating each entry embedding as an individual token, we concatenate the embeddings along each row to form feature vectors $\mathbf{F}_t \in \mathbb{R}^{N \times (N \cdot 8)}$. These are then projected via an MLP into topology tokens $\mathbf{F}_i \in \mathbb{R}^{N \times D}$ ($D=512$).
This design significantly reduces memory consumption during training while increasing the information density and semantic completeness of each token.
Moreover, this representation is highly interpretable: each token corresponds to one surface and explicitly encodes its adjacency relations to all other surfaces. 
Empirically, we observe that this row-wise tokenization achieves a substantially higher generation success rate than approaches that treat adjacency relations as independent scalar embeddings and rely on the network to implicitly learn structural dependencies.
Subsequently, the topology tokens and global conditioning signals are processed by a stack of DiT~\cite{ref/DiT} blocks to model global interactions. The conditioning signals consist of the number of surfaces and the embedding encoding of time t.
The Transformer output is projected via an MLP and reshaped into an $N \times N \times 8$ feature matrix. 
To ensure the topological consistency of the undirected adjacency graph, we enforce symmetry constraints by averaging the feature matrix with its transpose. 
Finally, an output MLP produces logits of shape $N \times N \times 3$, representing the categorical distribution of the adjacency states $\{0, 1, 2\}$.
The diffusion process is trained with the D3PM losses, combining a variational bound $\mathcal{L}_{vb}$ and an auxiliary loss on the reverse process: 
\begin{equation}
    \mathcal{L}_{\text{D3PM}}
= \mathcal{L}_{vb} + 0.02\cdot\mathbb{E}_{q(\mathbf{T})}\mathbb{E}_{q(\mathbf{T}_t|\mathbf{T})}
\left[
-\log{} p_{\theta_1}(\mathbf{T} | \mathbf{T}_t)
\right],
    \label{eq:topo-loss}
\end{equation}
where $q(\mathbf{T})$ is the topology distribution of the training dataset and $p_{\theta_1}(\mathbf{T}_0=\mathbf{T} | \mathbf{T}_t)$ can be derived from $p_{\theta_1}(\mathbf{T}_{t-1} | \mathbf{T}_t)$.

During inference, the noisy topology $\mathbf{T}_T$ is initialized as a set of mask tokens. We progressively denoise the representation into $\mathbf{T}_0$, conditioned on the diffusion time step and the surface count $N_i$. For the latter, we sample $N_i$ from the empirical distribution of the training dataset to ensure realistic complexity.
To regulate the diversity of adjacency generated, we introduce a time-dependent scaling factor $\eta$ on the output logits during inference.
By acting on the logarithmic outputs of D3PM, similar to the inverse temperature factor, the determinacy of the generation process can be effectively adjusted: its essence is to use the linear scaling of the logarithmic function to change the smoothness of the original probability distribution, reshape the entropy value of the Softmax distribution, and achieve a controllable balance between sample quality (Fidelity) and diversity (Diversity). 
A larger $\eta$ further amplifies the differences among the output logits, which sharpens the resulting probability distribution and consequently yields a higher degree of certainty.
\begin{equation}
\eta = 1 + m \left( 1 - \frac{t}{T} \right) + n,\quad m\geq 0,
\label{eq:scale}
\end{equation}
where $m, n$ govern the sharpness of the predicted probability distribution. 
By scheduling $\eta$ over sampling step $t$, we ensure high structural diversity during the initial diffusion steps ($\eta \rightarrow 1+n$) and gradually increasing prediction confidence toward the final steps ($\eta \rightarrow 1+m+n$) to ensure the validity of generated adjacency relationships.

For D3PM, we adopt a cosine noise scheduler, which ensures a smooth signal-to-noise ratio (SNR) evolution over timesteps, leading to more stable optimization and improved convergence for discrete state transitions.

\subsection{Topology-aware Surface Latent Encoding} \label{subsec:face_vae}
To encode the parametric surfaces $\mathbf{S}$ into a latent space, we employ a Variational AutoEncoder (VAE) with an encoder–decoder architecture primarily inspired by the UNet design in ~\cite{rombach22sd}, adopting an attention-augmented convolutional (AttnBlock-based) structure. In addition, we incorporate a central interaction mechanism after feature flattening, inspired by ~\cite{liu25hola}, to further enhance feature representation learning; unlike their surface-edge graph convolution, our design operates solely on a surface-level graph convolution.

\begin{figure}[h]
    \centering
    \includegraphics[width=1.0\linewidth]{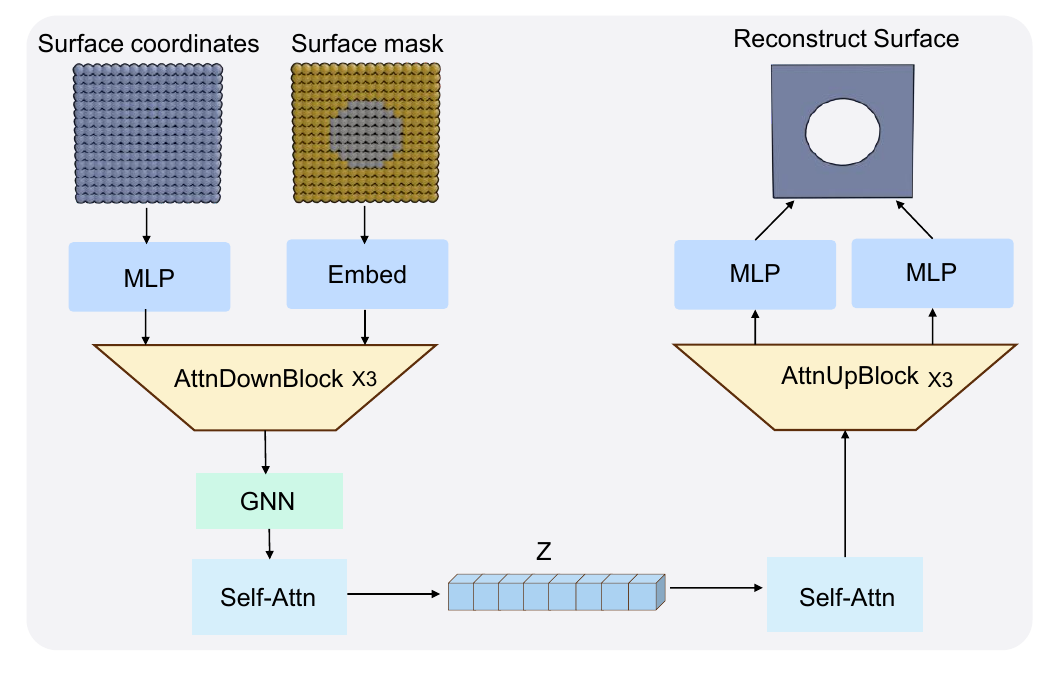}
    \caption{The architecture for the surface VAE network.}
    \label{fig:geom_vae}
\end{figure}

The surface VAE network is illustrated in Fig.~\ref{fig:geom_vae}.
The encoder takes as input $\mathbf{S}_{\text{mask}}$ and $\mathbf{S}_{\text{coord}}$, respectively.
Each input is mapped to feature tensors of size $\mathbb{R}^{N \times 4 \times 16 \times 16}$ and $\mathbb{R}^{N \times 16 \times 16 \times 16}$ via dedicated embedding layers, then concatenated channel-wise to form a joint $\mathbb{R}^{N \times 20 \times 16 \times 16}$ representation.
This representation is then processed by three convolutional encoder blocks with two downsampling operations, yielding feature maps of size $\mathbb{R}^{N \times 8 \times 4 \times 4}$.
The spatial features are further flattened into 128-dimensional feature vectors for each surface.
On top of the flattened features, two graph neural network layers~\cite{chen20gcn} are applied, where surface adjacency is used to guide message passing and explicitly model topological relationships among surfaces, followed by two self-attention layers to further aggregate global contextual information.
The encoder outputs the Gaussian mean and variance for latent sampling $\mathbf{\mu}, \mathbf{\sigma} \in \mathbb{R}^{N \times D}$. We denote the encoded latent with $\mathbf{Z}$.

The decoder takes as input a sampled surface latent $\hat{\mathbf{Z}} \sim \mathcal{N}(\mathbf{\mu}, \mathbf{\sigma}^2) \in \mathbb{R}^{N \times D}$.
The sampled latents are first processed by two self-attention layers and then reshaped into spatial feature maps of size $N \times 8 \times 4 \times 4$.
Subsequently, three convolutional decoder blocks with two upsampling operations are applied.
The decoder then branches into two output heads of the surface coordinate map $\hat{\mathbf{S}}_{\text{coord}} \in \mathbb{R}^{N \times 3 \times 16 \times 16}$ and a binary validity mask $\hat{\mathbf{S}}_{\text{mask}} \in \mathbb{R}^{N \times 1 \times 16 \times 16}$.

The overall objective for VAE optimization $\mathcal{L}_\text{VAE}$ combines a surface reconstruction loss $\mathcal{L}_\text{rec}$ between $\hat{\mathbf{S}}$ and $\mathbf{S}$, a Kullback-Leibler (KL) divergence $\mathcal{L}_\text{KL}$, and a Maximum Mean Discrepancy (MMD) regularization $\mathcal{L}_\text{MMD}$ that aligns the latent distribution with the standard Normal distribution. 
The implementation of $\mathcal{L}_\text{KL}$ and $\mathcal{L}_{\mathrm{MMD}}$ follows ~\cite{zhao19infovae}, while the reconstructio loss is defined as $\mathcal{L}_\text{rec}
=
\|\mathbf{S}_{\text{coord}} - \hat{\mathbf{S}}_{\text{coord}}\|_2
+
0.001\cdot\mathrm{BCE}(\mathbf{S}_{\text{mask}}, \hat{\mathbf{S}}_{\text{mask}})$.

The loss definitions are as follows: 
\begin{equation}
    \mathcal{L}_\text{VAE} = w_1\mathcal{L}_\text{VAE-rec} + w_2\mathcal{L}_\text{reg} + w_3\mathcal{L}_\text{MMD},
\end{equation}
where $\text{BCE}(\cdot)$ denotes the binary cross-entropy loss and balancing weights are set as $w_1 = 1, w_2 = 10^{-7}, w_3 = 10^{-4}$.

\subsection{Topology-Conditioned Surface Diffusion}\label{subsec:face-diff}
The surface diffusion model learns the distribution of surface latents $\mathbf{Z}$.
%
%
\begin{figure}
    \centering
    \includegraphics[width=\linewidth]{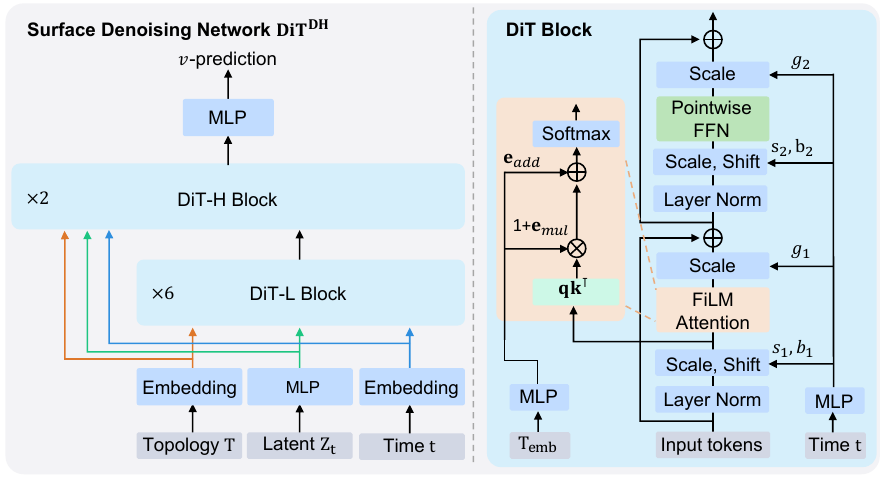}

    \caption{The architecture for the surface denoising network. Left: overall network. Right: DiT block detail in $\text{DiT}^{\text{DH}}$, DiT-L and DiT-H differ only in the feature dimension.}
    \label{fig:geom_diffusion}
\end{figure}
The Geometry denoising network is illustrated in Fig.~\ref{fig:geom_diffusion}.
First, the network embeds the input features. 
The noisy latent features $\mathbf{Z}_t$ are encoded into $D$-dimensional input tokens using an MLP.
The timestep $t$ is encoded using a sinusoidal positional embedding followed by an MLP to channel-wise modulation parameters $s_i, b_i, g_i$.
The generated topology $\mathbf{T}$ is first embedded as $\mathbf{T}_{emb}$ and then fed to an MLP, outputting attention modulation parameters $\mathbf{e_\mathrm{mul}, e_\mathrm{add}}$. 
Subsequently, embedded inputs are fed to a DiT-based network~\cite{ref/DiT}.
Our denoising network design is inspired by the $\mathrm{DiT}^{\mathrm{DH}}$ architecture proposed in ~\cite{zheng25rae} and consists of six DiT-L blocks and two DiT-H blocks. 
This design is introduced to improve training efficiency by reducing parameters and computational cost while maintaining high-quality generation. 
The first six DiT-L blocks use a smaller feature dimension of 384 to reduce overhead, and their outputs are propagated to subsequent DiT-H blocks with higher feature dimensions of 768, enabling the network to progressively refine geometric details while achieving a more parameter-efficient design without compromising geometric fidelity.
Within each DiT-L and DiT-H block, standard self-attention is replaced by FiLM-style modulated attention~\cite{perez18film} in order to incorporate topology for consistency between surface topology and surface geometry.
Since the topology condition does not use positional encoding and each face only attends to its adjacent faces, the topology representation is permutation-invariant. Consequently, permuting the rows and columns of the topology matrix results in an equivalent topology during topology generation.
Denoting the query and key of head $i$ as $\mathbf{q}$ and $\mathbf{k}$, the FiLM modulated attention map is computed as
\begin{equation}
\text{Attn} = 
\text{softmax}\Bigg(
(1 + \mathbf{e_\mathrm{mul}}) \odot \frac{\mathbf{q} \mathbf{k}^\top}{\sqrt{d}} + \mathbf{e_\mathrm{add}}
\Bigg),
\label{eq:topo-modulated-attn}
\end{equation}
where $\odot$ denotes element-wise multiplication. This mechanism modulates the attention scores with topology information, enforcing topology-aware feature interactions.
Specifically, for a multi-head self-attention with $H$ heads, the topology embedding $\mathbf{T}_\text{emb}$ is first mapped via an MLP into a modulation tensor of size $N \times N \times 2H$. This tensor is then chunked along the last dimension to produce the per-head scale $\mathbf{s} \in \mathbb{R}^{N \times N \times H}$ and bias $\mathbf{b} \in \mathbb{R}^{N \times N \times H}$ for each attention head.  

The overall diffusion process is trained with a velocity prediction objective~\cite{ref/v-prediction}:
\begin{equation}
\mathcal{L}_{v-\text{pred}}
=
\mathbb{E}_{\mathbf{Z}, t, \mathbf{T}}\left [\left\|
\mathbf{v}_{\theta_2}(\mathbf{Z}_t, t, \mathbf{T})
-
v_t
\right\|_2^2\right],
\end{equation}
where the target velocity $v_t$ is defined as
\begin{equation}
v_t =
\sqrt{\bar{\alpha}_t}\,\boldsymbol{\epsilon}
-
\sqrt{1-\bar{\alpha}_t}\,\mathbf{Z}_0.
\end{equation}
We adopt the velocity parameterization instead of directly predicting the noise, as it provides improved training stability and better empirical performance. 


\subsection{Post-processing}\label{subsec:pp}
\begin{figure}[h]
    \centering
    \includegraphics[width=\linewidth]{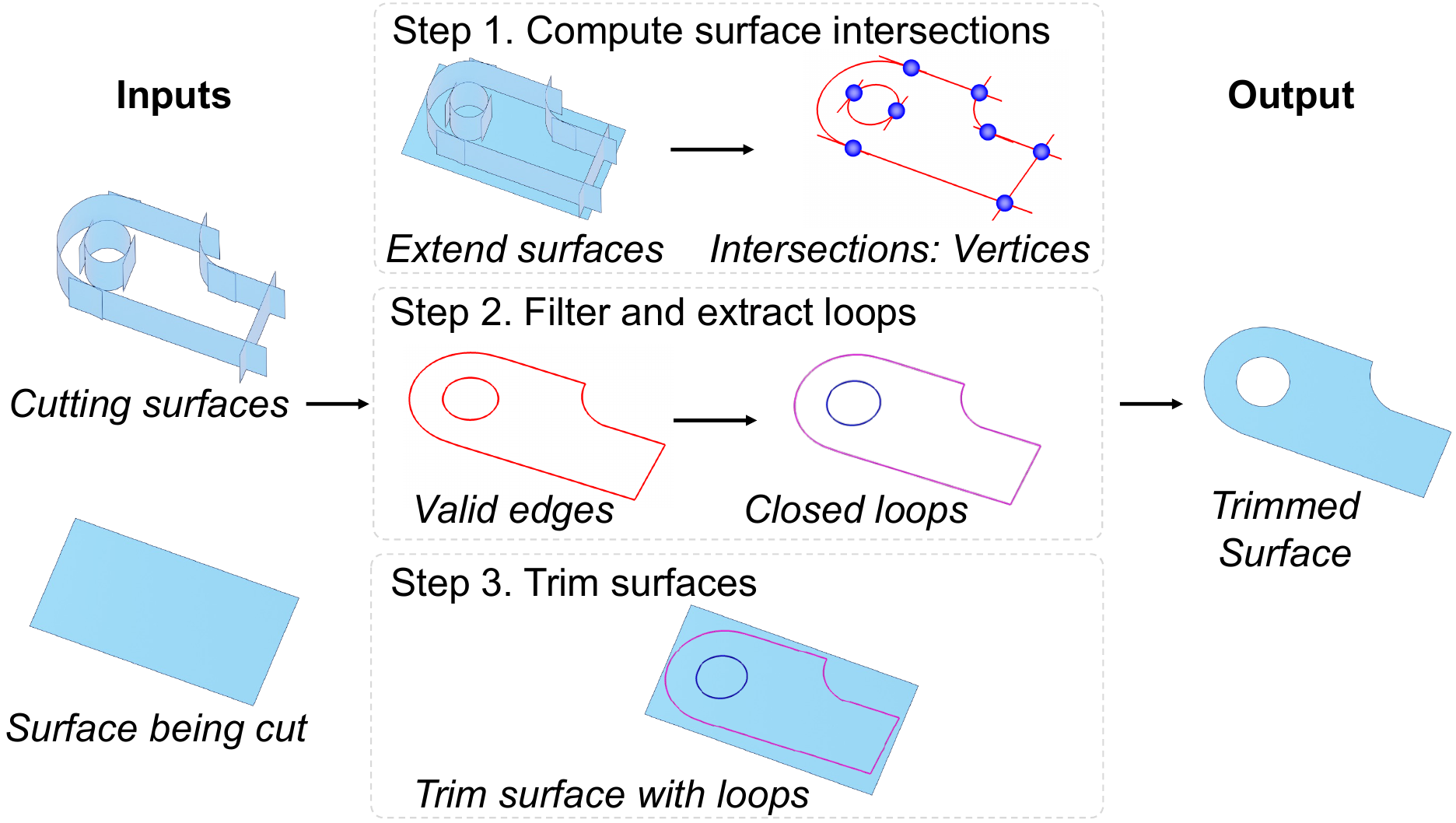}
    \caption{An illustration of the surface trimming algorithm. Left: The algorithm has multiple cutting surfaces and a target surface as input. Middle: The algorithm proceeds in three steps. Right: The output is the trimmed target surface.}
    \label{fig:postprocess}
\end{figure}

Our post-processing pipeline takes as input the generated surface adjacency and surface geometry and outputs the post-processed B-rep shape. 
We first identify the surface type for each surface by fitting its point set with candidate analytic surfaces, including planes, cylinders, cones, tori, and spheres, and selecting the type that yields the minimum fitting error. 
If the bidirectional fitting error is below a threshold of $0.025$, the face is replaced by its corresponding analytic surface representation; otherwise, a B-spline surface is constructed from the predicted points. 
The fitting parameters are specific to the surface type, for example, a cylinder with parameters of radius, height, and orientation angle, and a cone with parameters of the slope and angle.
Next, we trim each surface with a surface-cutting module based on surface adjacency, as illustrated in Fig.~\ref{fig:postprocess}. The input consists of a surface to be cut and multiple cutting surfaces. 
All surfaces are first extended along the UV direction to ensure robust intersection, after which pairwise surface intersections are computed to obtain candidate intersection curves. 
Intersections among these curves further yield vertex positions. 
Based on the extracted vertices, curve segments are filtered such that only those with two valid intersection endpoints are retained. 
Subsequently, a minimum-cycle detection algorithm~\cite{kavitha2007minimum} is applied to partition the retained curve segments into closed loops. This process ensures that the extracted loops correspond to the tightest bounded regions without internal boundaries, thereby preventing the formation of overlapping or composite loops. 
Finally, Boolean operations are used to handle cases where one loop encloses another, ensuring that all trimmed faces are non-overlapping. Faces are then filtered by enforcing that every edge has degree 2. 
After surface trimming, all surfaces are assembled into a Brep model based on their mutual adjacency.

\begin{figure}[h]
    \centering
    \includegraphics[width=\linewidth]{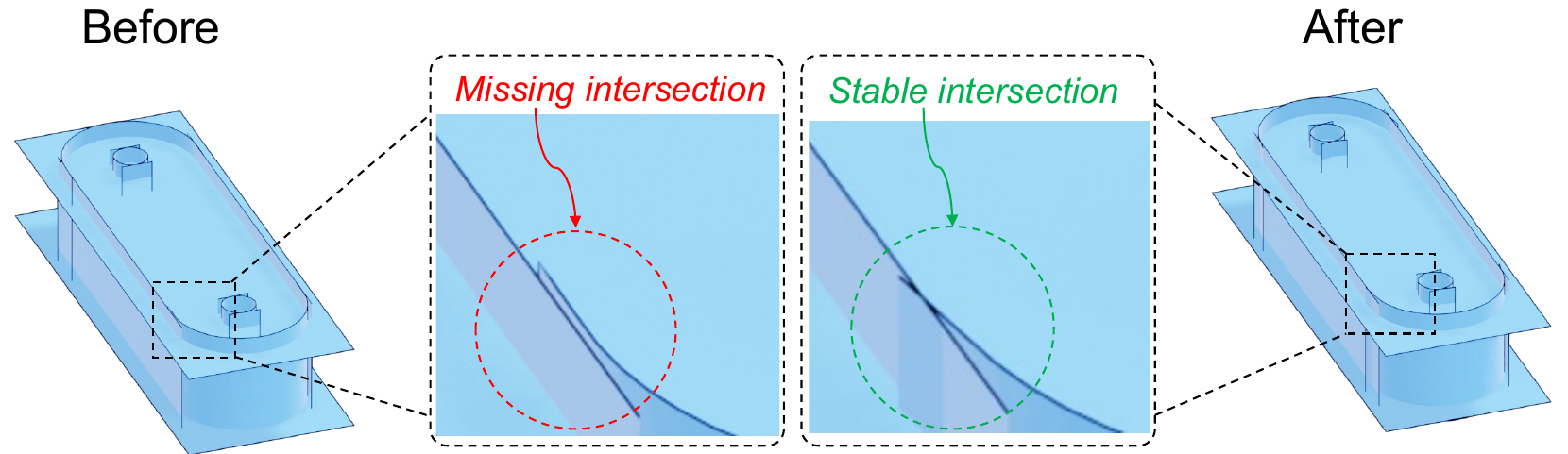}
    \caption{An illustration for surface trimming for near-tangent cases. Left: Near-tangent surfaces have a missing intersection. Right: After angle adjustments, most problematic cases can be resolved.}
    \label{fig:change}
\end{figure}

Though the surface-cutting module can handle most cases, it may fail to obtain intersection curves when the intersected surfaces are nearly tangent, as illustrated in Fig.~\ref{fig:change}. This problem is also noted in the appendix of BRepDiff~\cite{lee25brepdiff}.
To address this issue, we slightly adjust the angular parameters of circular surfaces, which increases the likelihood of surface intersections when a closed edge loop cannot be formed. We adopt an iterative dynamic trimming strategy that only adjusts the angles of failed surfaces.
This approach (with at most three retries) effectively overcomes tangency issues.

\begin{figure}[h]
    \centering
    \includegraphics[width=\linewidth]{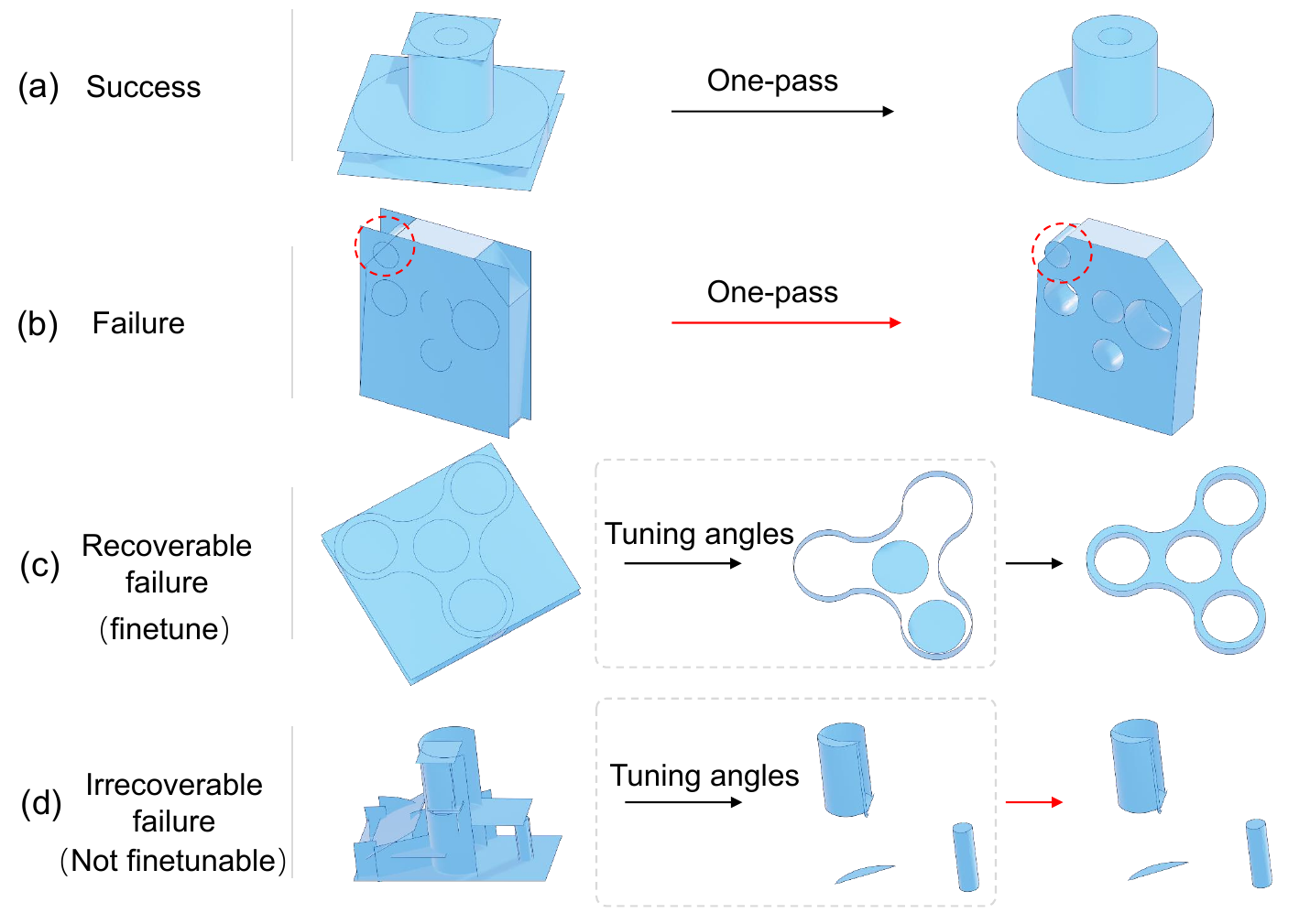}
    \caption{Typical examples for post-processing results.
    Case (a) corresponds to a successful single-pass post-processing result that requires no further adjustment, which accounts for the majority of cases. 
    Case (b) represents a rare failure case where the model is geometrically incorrect from a human perspective and cannot be resolved by post-processing.
    Case (c) corresponds to a typical failure case, where an initial post-processing failure is resolved by adjusting the fitting parameters of the failed surfaces in subsequent retries. 
    Case (d) corresponds to a failure case, which may be caused either by topological inconsistencies or by overly complex geometry leading to geometric generation errors and post-processing failures. In this case, surfaces cannot form a valid solid regardless of the number of retries.
    }
    \label{fig:postprocess_type}
\end{figure}

To further demonstrate the post-processing effects, Fig.~\ref{fig:postprocess_type} illustrates several typical examples. 

\section{Experiments} \label{sec:exp}

\subsection{Experimental Setups}\label{subsec:setup}
\noindent \textbf{Datasets.} We evaluate our method on the DeepCAD~\cite{wu21deepcad}, ABC~\cite{koch19abc} datasets, and furniture dataset collected from the OnShape dataset repository, adopting the preprocessing protocols and partition splits established by BrepGen~\cite{xu24brepgen}:
the DeepCAD subset (DeepCAD-0-30) is restricted to shapes with a maximum of $30$ surfaces and $20$ edges per surface, while the ABC dataset (ABC-0-50) and furniture dataset (Furniture-0-50) are filtered to include models with up to $50$ surfaces and $30$ edges per surface.
Furthermore, we introduce \emph{DeepCAD-7-30}, a subset specifically curated by selecting models with $7$ to $30$ faces. We use this dataset for ablation experiments.
The number of training samples is 83,755 for DeepCAD-0-30, 63,418 for DeepCAD-7-30, 293,457 for ABC-0-50, and 1,341 for Furniture-0-50.

\head{Training Details.} 
All models in our framework, including the VAE, topology diffusion model, and geometry diffusion model, are trained on four NVIDIA RTX 3090 GPUs. We use the Adam optimizer for all components, with a learning rate of $5\times10^{-4}$, momentum parameters $\beta = (0.90, 0.95)$, and a weight decay of $1\times10^{-6}$. A cosine learning rate scheduler with warmup is applied throughout training.
The VAE is trained for $1000$ epochs with a batch size of $512$, during which random rotations of the entire shape are applied as data augmentation.
The topology diffusion model is trained with $200$ denoising steps and a batch size of $3072$ for $6000$ epochs. 
The geometry diffusion model is trained with $1000$ denoising steps and a batch size of $1024$ for $6000$ epochs. 

\head{Inference.} 
During inference, the number of surfaces is first sampled according to the empirical distribution observed in the original training dataset ~\cite{you2018graphrnn}. Specifically, we estimate the proportion of shapes corresponding to each surface count and randomly sample the surface number following this distribution.
Then, we generate surface topology by denoising an initial topology with all [MASK] states, conditioned on the sampled surface count.
Surface geometry is generated by progressively denoising from Gaussian noise, conditioned on the generated topology.
The final solid is obtained through our post-processing procedure, which relies solely on the predicted topology and input surfaces. By extending adjacent surfaces and computing their intersections, a B-rep shape is formed. 
The inference configuration is 1 × RTX 3090 GPU with batch size of 4096, and post-processing on CPU.

\head{Metrics.} 
We follow the standard evaluation metrics for 3D generation used in prior work~\cite{xu24brepgen}, and additionally introduce several complementary metrics. All metrics are listed as follows.

\begin{itemize}[leftmargin=5mm]
    \item Coverage (COV.) $\uparrow$ measures how well the generated set covers the diversity of the ground-truth dataset.
    \item Minimum Matching Distance (MMD.) $\downarrow$ measures the overall quality of generated shapes by computing, for each ground-truth shape, the distance to its closest generated shape and averaging over all ground-truth samples.
    \item Jensen–Shannon Divergence (JSD.) $\downarrow$ measures distribution similarity between real and generated shapes based on statistical properties.
    \item Validity (Valid.) $\uparrow$ measures the proportion of generated CAD models that are a valid B-rep shape with watertight surfaces.
    \item Cyclomatic Complexity (CC) $\uparrow$ measures the structural complexity of a B-Rep model by counting independent cycles in its wireframe graph representation, following Contero et al.~\cite{contero2023cadcomplexity}. 
    \item Vision-Language Model Complexity (VLM-C)$\uparrow$ measures the reasoning difficulty for a vision-language model to interpret a generated shape, reflecting the structural complexity of the underlying geometry. 
    \item Vision-Language Model Quality (VLM-Q) $\uparrow$ evaluates the perceived geometric and visual quality of generated shapes using a vision-language model, reflecting surface smoothness, structural plausibility, and overall fidelity. 
\end{itemize}

\begin{table*}[h]
    \centering
    \caption{Quantitative comparison of unconditioned B-rep generation on the DeepCAD-0-30 and ABC-0-50 datasets. We report unconditional generation results. Please note that MMD and JSD values are multiplied by 100, while COV and validity metrics are expressed as percentages. All experiments are run 10 times and are reported on average. In addition, the maximum value of VLM metrics is 10.
    }
    \label{tab:comparison_uncondition}
    \begin{tabular}{c|ccccccc}
    \toprule
    {DeepCAD-0-30}  & COV(\%) $\uparrow$ & MMD($10^{-2}$) $\downarrow$ & JSD($10^{-2}$) $\downarrow$ & Valid(\%) $\uparrow$ & CC(\%) $\uparrow$ & VLM-C(\%) $\uparrow$ & VLM-Q(\%) $\uparrow$ \\ \midrule
    DeepCAD                 & 71.1  &1.19  &1.70  & 63.1 & 8.8  & 4.2 & 7.6\\ 
    BrepGen                 & 71.7  &1.34  &1.95  & 50.5 & 9.1  & 4.7 & 6.5\\ 
    DTGBrepGen              & 73.9  &1.10  &1.32  & 75.6 & 11.4 & 5.7 & \textbf{8.8}\\
    BrepDiff                & 73.5  &1.14  &1.41  & 59.3 & 9.4  & 5.4 & 6.9\\
    \hline
    Ours   & \textbf{77.9}& \textbf{1.03} & \textbf{1.07} &  \textbf{85.6} & \textbf{12.6} & \textbf{6.6} & 8.7 \\
    \bottomrule
    \addlinespace[5pt]
    \toprule
    {ABC-0-50}  & COV(\%) $\uparrow$ & MMD($10^{-2}$) $\downarrow$ & JSD($10^{-2}$) $\downarrow$ & Valid(\%) $\uparrow$ & CC(\%) $\uparrow$ & VLM-C(\%) $\uparrow$ & VLM-Q(\%) $\uparrow$ \\ \midrule
    BrepGen                 & 69.1 & 1.46 & 2.08 & 41.4 & 11.0  & 5.4 & 5.9\\ 
    DTGBrepGen              & 72.3 & 1.23 & 1.54 & 52.9 & 13.2  & 6.3 & 8.2\\
    BrepDiff                & 71.9 & 1.29 & 1.56 & 30.1 & 11.9  & 6.1 & 6.5\\
    \hline
    Ours & \textbf{74.3} & \textbf{1.17} & \textbf{1.13} & \textbf{61.4} & \textbf{15.1}  & \textbf{7.2} & \textbf{8.5}\\
    \bottomrule
    \end{tabular}
\end{table*}

\begin{table}[h]
    \centering
    \caption{Quantitative comparison of unconditioned B-rep generation on the DeepCAD-0-30 and ABC-0-50 datasets without post-processing.
    }
    \setlength{\tabcolsep}{1mm}
    \label{tab:comparison_uncondition_no_postprocess}
    \begin{tabular}{c|ccc}
    \toprule
    {DeepCAD-0-30}  & COV*(\%) $\uparrow$ & MMD*($10^{-2}$) $\downarrow$ & JSD*($10^{-2}$) $\downarrow$  \\ \midrule
    BrepGen                 & 73.7  &1.25  &1.58  \\ 
    DTGBrepGen              & 77.6  &0.88  &1.21  \\
    BrepDiff                & 78.2  &0.86  &1.18  \\
    \hline
    Ours               & \textbf{80.8}& \textbf{0.80} & \textbf{1.04} \\
    \bottomrule
    \addlinespace[5pt]
    \toprule
    {ABC-0-50}  & COV*(\%) $\uparrow$ & MMD*($10^{-2}$) $\downarrow$ & JSD*($10^{-2}$) $\downarrow$ \\ \midrule
    BrepGen                 & 72.5 & 1.35 & 1.68  \\ 
    DTGBrepGen              & 75.1 & 1.12 & 1.36  \\
    BrepDiff                & 78.2 & 1.15 & 1.31  \\
    \hline
    Ours               & \textbf{80.1} & \textbf{0.97} & \textbf{1.10} \\
    \bottomrule
    \end{tabular}
\end{table}

For evaluation, we randomly generate 3000 shapes to compute the metrics, and use * to represent the indicators for results without post-processing, such as (COV*), (MMD*), (JSD*). 
Additionally, we adopt the Qwen3-VL-8B~\cite{bai2023qwenvl} for VLM-based evaluation. 
To validate the effectiveness of the VLM-based metrics, we conducted a user study with 20 college-educated participants. Users were asked to evaluate the complexity and regularity of 110 B-rep shapes. We then compared the VLM predictions with human judgments, achieving a consistency rate of 90.4\% for the quality scores and 86.5\% for the complexity scores. 

\subsection{Comparison Study}
We evaluate our approach against DeepCAD~\cite{wu21deepcad} and several recent diffusion-based methods~\cite{li25dtg,xu24brepgen,lee25brepdiff}.
We did not compare with HoLa~\cite{liu25hola} as there is no available code for evaluation. 
All methods were evaluated using their publicly available code and checkpoints retrained by us on the standardized training sets of DeepCAD-0-30 and ABC-0-50.
This section presents extensive comparisons in unconditional generation, topology generation, and running performance.

\begin{figure}[h]
    \centering
    \includegraphics[width=1.0\linewidth]{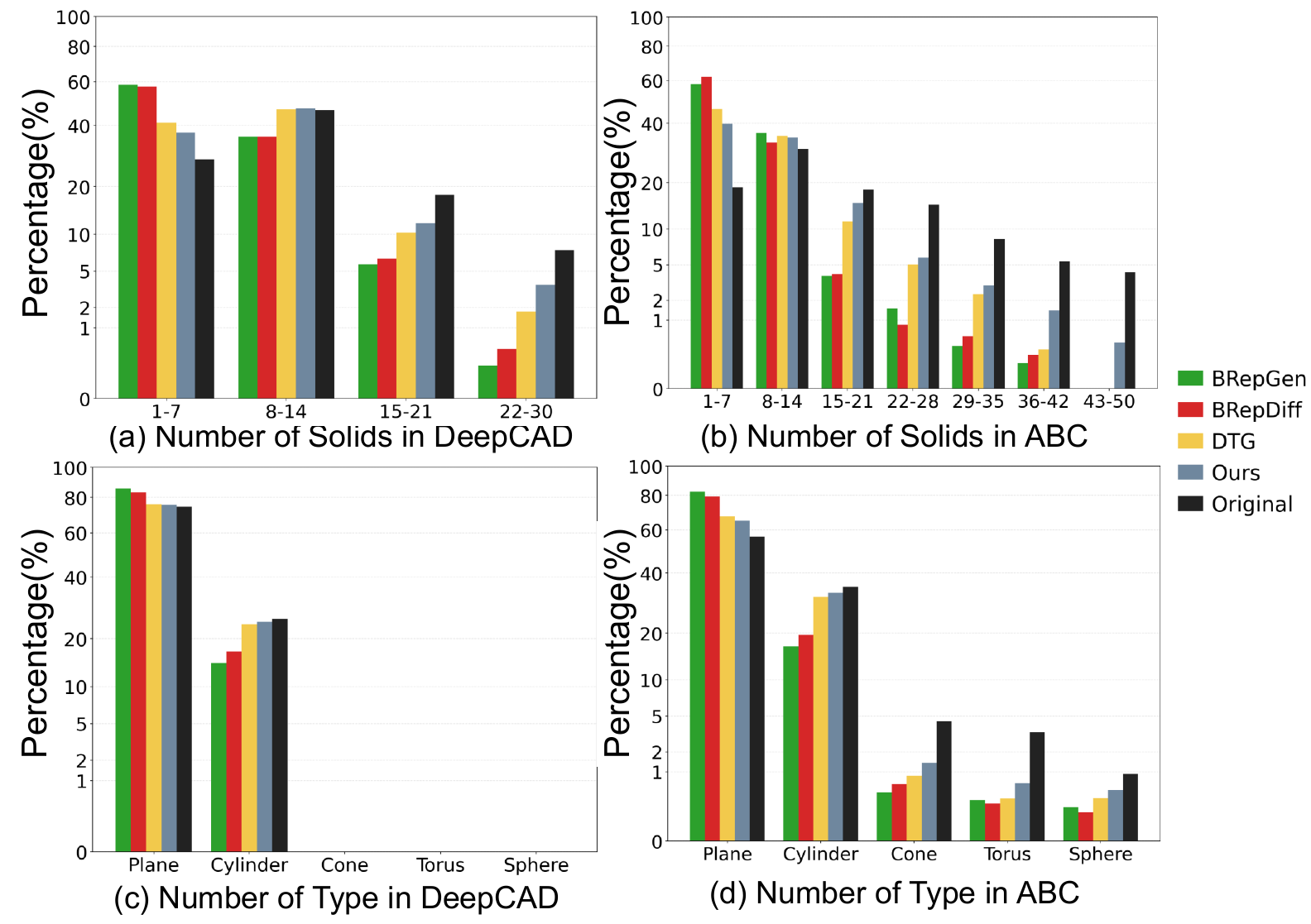}
    \caption{Distribution of the dataset and generated results of different methods. Ours is closest to the ground truth distribution.}
    \label{fig:res_ratio}
\end{figure}

\head{Unconditional Generation.}
We analyze the distribution of surface counts and surface types of the generated results, as shown in Fig.~\ref{fig:res_ratio}. Surface counts are directly computed, while surface types are estimated via patch fitting. 
For surface counts, as shown in (a) and (b), our method better matches the training datasets and achieves higher average surface numbers on both the DeepCAD and the ABC dataset. This is because our model produces fewer samples with low surface counts and more samples with high surface counts. In particular, we significantly outperform prior methods in the ranges of 23–30 surfaces on DeepCAD and 43–50 surfaces on ABC.
For surface types, as shown in (c), curved surfaces dominate in our generated results, with proportions close to DTGBrepGen. As illustrated in (d), our method maintains higher proportions across various curved surface types, indicating that our generated shapes better match the training dataset distribution in terms of structural complexity.

\begin{figure*}
    \centering
    \includegraphics[width=1.0\linewidth]{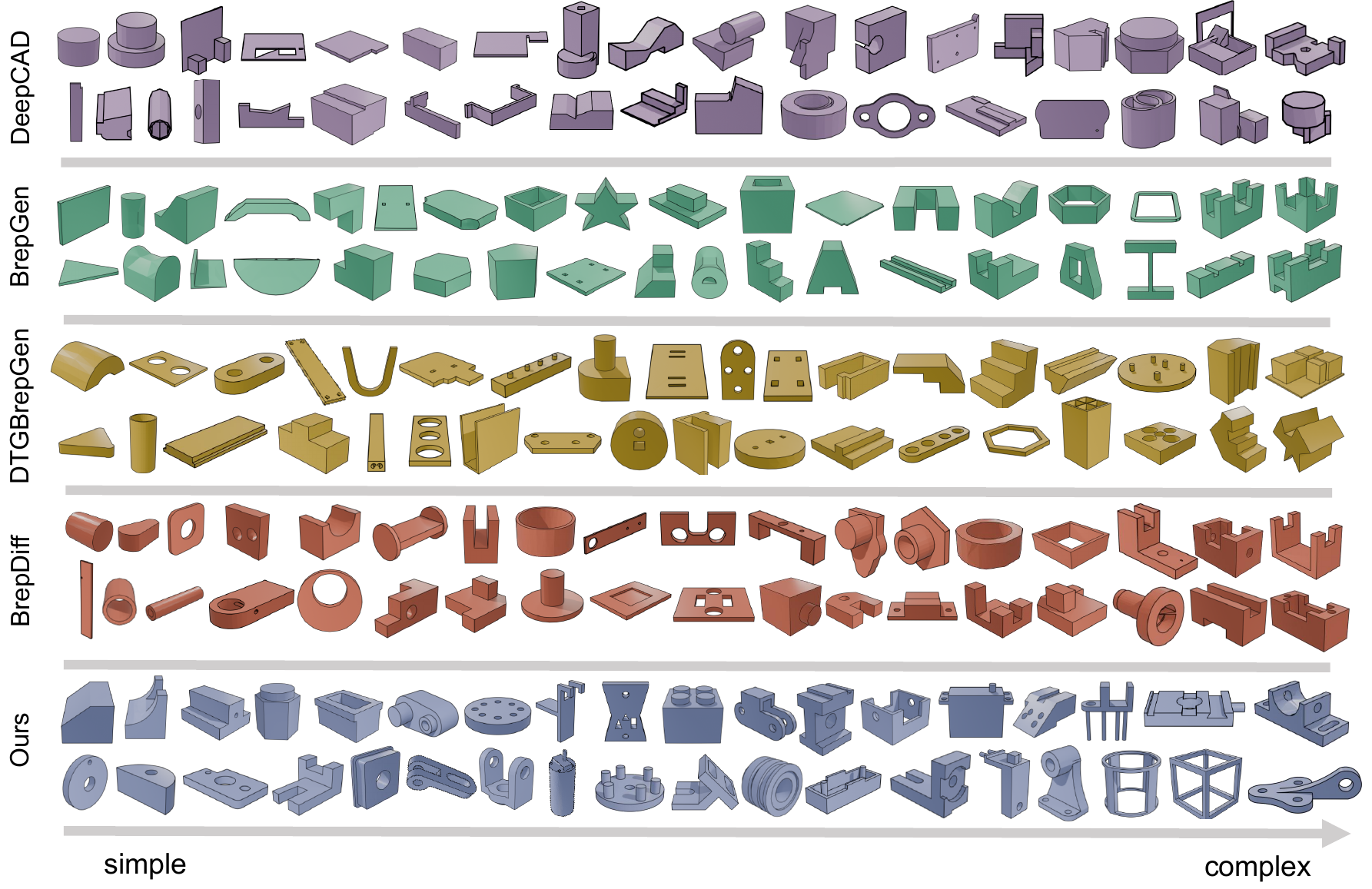}
    \caption{Visual comparisons of unconditional generation on the DeepCAD-0-30 dataset. From left to right, the complexity of the generated B-rep increases.}
    \label{fig:exp_deepcad}
\end{figure*}

\begin{figure*}
    \centering
    \includegraphics[width=1.0\linewidth]{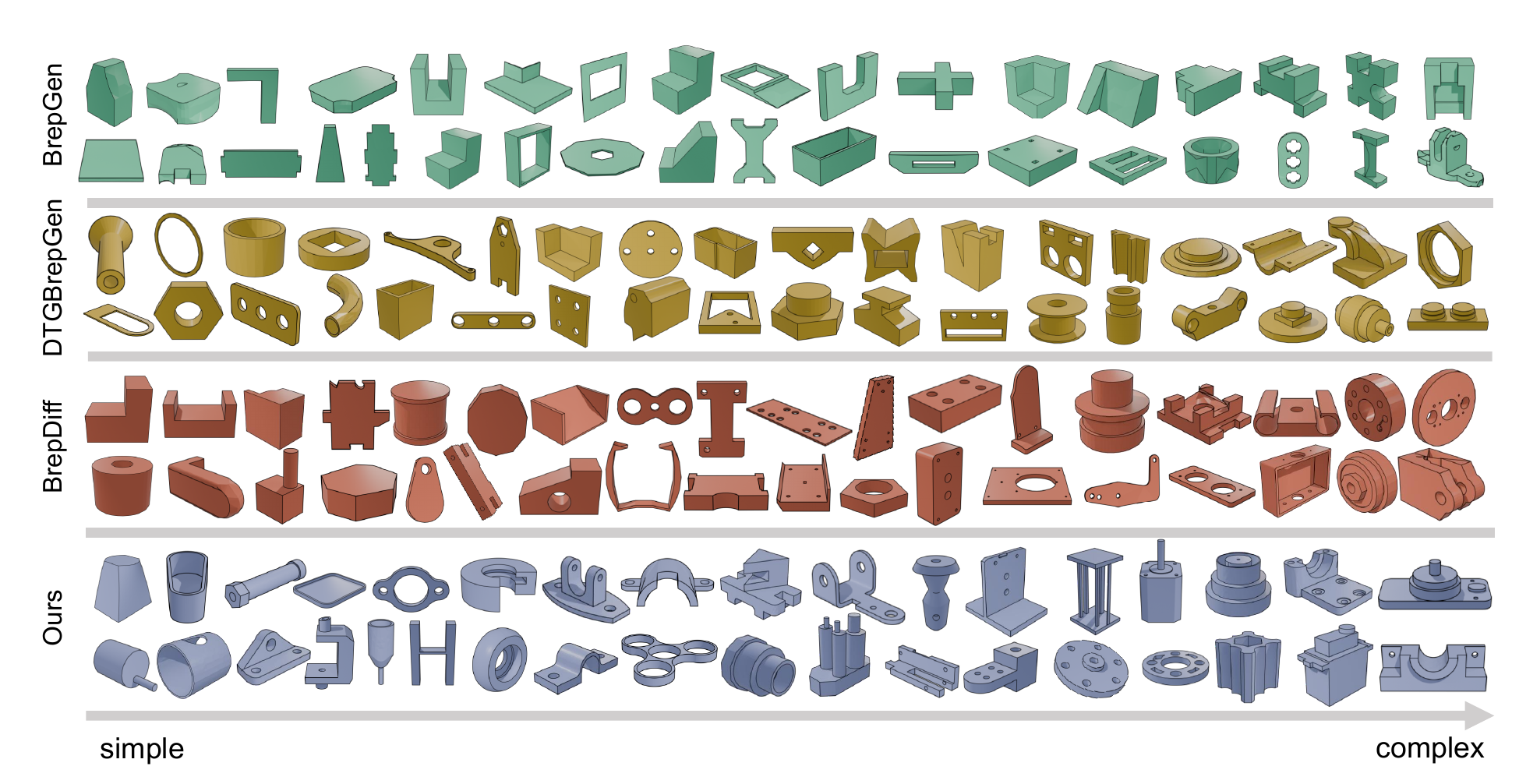}
    \caption{Visual comparisons of unconditional generation on the ABC-0-50 dataset. From left to right, the complexity of the generated B-rep increases.}
    \label{fig:exp_abc}
\end{figure*}

\begin{figure*}
    \centering
    \includegraphics[width=1.0\linewidth]{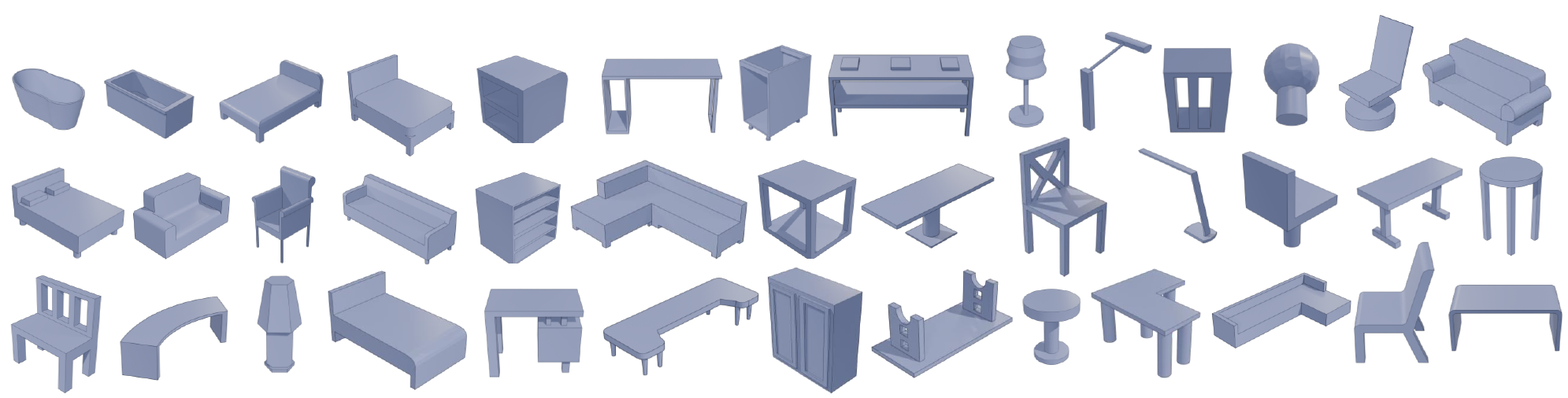}
    \caption{Visual comparisons of unconditional generation on the Furniture-0-50 dataset.}
    \label{fig:exp_furniture}
\end{figure*}


Tab.~\ref{tab:comparison_uncondition} presents the quantitative results for unconditional B-rep generation on the DeepCAD-0-30 and ABC-0-50 datasets. We compare the results with post-processing enabled for all competing methods.
Our method achieves significant improvements across metrics on both datasets.
Our method achieves the best performance in terms of COV, MMD, JSD, and validity, outperforming all other baselines. In addition, it obtains the highest CC values, indicating that our generated results are structurally more complex. 
In the VLM-based assessment, our method leads for ABC and is second for DeepCAD.
Visual results in Fig.~\ref{fig:exp_deepcad} and Fig.~\ref{fig:exp_abc} demonstrate that our method generates significantly more diverse geometric structures. In particular, our results exhibit complex hole patterns, rich combinations of curved surfaces, and repetitive structures such as cylindrical holes and pillars.
In contrast, competing methods tend to produce geometries dominated by flat surfaces, with only simple curved regions and limited hole structures.
Moreover, the results produced by BrepGen and BrepDiff frequently exhibit cracks or irregular meshes, as highlighted, whereas such artifacts are largely absent in our outputs.

Tab.~\ref{tab:comparison_uncondition_no_postprocess} reports the results without post-processing, where metrics are computed directly from the raw patches output by the network. 
This provides a coarse evaluation of the model’s intrinsic generation capability.
For consistency, the reference data are obtained via UV sampling followed by B-spline fitting and contains uncut points (see the left and right figures in Fig.~\ref{fig:postprocess_type} for an example). Because there is no post-processing and only the results generated by the network are considered, there is no validity metric available.
As shown in the results, our method achieves the best performance across all metrics, demonstrating strong intrinsic generation capability and better alignment with the underlying data distribution.
Note that the metrics in Tab.~\ref{tab:comparison_uncondition} and Tab.~\ref{tab:comparison_uncondition_no_postprocess} are not directly comparable, as the GT reference is different (fitted B-spline patches vs GT B-rep surfaces). 

\head{Topology Generation.}
We further evaluate our approach against the autoregressive surface topology generation of DTGBrepGen, specifically on topological generation. 
\begin{table}[h]
    \centering
    \caption{Comparison of the rationality of generated topology.}
    \label{tab:exp_topo_quality}
    \vspace{-3mm}
    \resizebox{\linewidth}{!}{
    \begin{tabular}{c|cc|cc}
    \toprule
     & \multicolumn{2}{c|}{DeepCAD-0-30} & \multicolumn{2}{c}{ABC-0-50} \\
     &  Ours  & DTGBrepGen &  Ours  & DTGBrepGen \\
    \midrule
    Valid-topo (\%) $\uparrow$          & \textbf{99.8}  & 97.5 & \textbf{96.5} & 86.8 \\ 
    Uniqueness (\%) $\uparrow$          & \textbf{36.2}  & 19.8 & \textbf{46.5} & 42.5 \\ 
    \bottomrule
    \end{tabular}
    }
\end{table}
In this assessment, \emph{valid-topo}$\uparrow$ denotes whether the generated topology is fully connected and symmetric, while \emph{uniqueness}$\uparrow$ quantifies the proportion of non-isomorphic structures within the generated set. 
Two topologies are defined as isomorphic if a permutation of elements exists that maps one to the other. 
To identify duplicates, we utilize the BLISS method~\cite{bliss07,bliss11} to reduce two topologies to a canonical form, subsequently checking for adjacency matches. For both methods, metrics are computed based on 1,000 randomly generated samples.

As shown in Tab.~\ref{tab:exp_topo_quality}, our method achieves a topology success rate exceeding $95\%$ on both the DeepCAD and ABC datasets.
On the ABC-0-50 dataset, our approach improves the success rate by approximately 9\% compared to DTGBrepGen.
Furthermore, the volume of non-isomorphic topologies generated by our model is nearly double that of the baseline on DeepCAD-0-30 and remains superior to that of the baseline on ABC-0-50. 
These results indicate that our framework significantly outperforms autoregressive topology generation in both structural validity and diversity.

\head{Post-processing Comparison.}
Tab.~\ref{tab:only-post-processing} presents a comparison between our post-processing method and that of BrepDiff on the DeepCAD-0-30 dataset. Since both approaches are surface-based generation methods, the results show that incorporating topology-aware post-processing outperforms methods without such geometric and topological guidance. Moreover, as the number of surfaces increases, particularly in the range of 21 to 30, our method demonstrates a significant advantage, achieving approximately three times higher rationality than BrepDiff.

\begin{table}[h]
    \centering
    \setlength{\tabcolsep}{16pt} 
    \caption{Comparing the validity of using raw data only after post-processing in DeepCAD-0-30 dataset, using different ranges of surface numbers.}
    \label{tab:only-post-processing}
    \vspace{-3mm}
    \resizebox{\linewidth}{!}{
    \begin{tabular}{c|cccc}
    \toprule
      &  0-10  & 11-20 &  21-30 & all   \\
    \midrule
    Ours             & \textbf{99.2}  & \textbf{87.3} & \textbf{74.5} & \textbf{95.6} \\ 
    BrepDiff         & 92.6 & 62.3  & 26.5 & 80.5 \\ 
    \bottomrule
    \end{tabular}
    }
\end{table}

We compare different post-processing methods in Tab.~\ref{tab:Comparison-of-post-processing}, highlighting that all methods require post-processing of comparable complexity. Only our method and DTGBrepGen employ primitive fitting, and our approach demonstrates more stable and robust fitting results than DTGBrepGen.
Both BrepGen and DTGBrepGen explicitly generate edges, while BrepDiff obtains edges during post-processing by performing mesh intersection followed by fitting. HOLA infers edge existence and intersections via a VAE-based prediction of intersecting surfaces. In contrast, our method fits surfaces into OCC geometry and derives edges through extended intersection operations.
Geometry refinement is commonly involved in all post-processing pipelines. BrepGen and DTGBrepGen apply 200 steps of gradient descent to better fit edges to surfaces, along with three thresholding attempts to prevent errors when converting patch points to B-splines. BrepDiff only uses the three thresholding attempts. 
Meanwhile, to maintain consistency, our method performs at most three post-processing steps, where each step only adjusts the fitting angles of failed surfaces, while planar surfaces remain unchanged.

\begin{table}[h]
    \centering
    \caption{Comparison of post-processing methods.}
    \label{tab:Comparison-of-post-processing}
    \vspace{-3mm}
    \resizebox{\linewidth}{!}{
    \begin{tabular}{c|ccc}
    \toprule
      &  Primitive Fitting  & Edge extraction & Geometry tuning  \\
    \midrule
    BrepGen                 & \texttimes  &Generated & 200-step Optimization and Threshold  \\ 
    DTGBrepGen              & \checkmark  &Generated & 200-step Optimization and Threshold  \\
    BrepDiff                & \texttimes  &Intersection(Mesh)  &Threshold  \\
    Hola                    & \texttimes  &Intersection(VAE)   & N/A \\
    \hline
    Ours   & \checkmark & Intersection(Geom)  & Adjust fitting angle \\       
    \bottomrule
    \end{tabular}
    }
\end{table}

\head{Runtime Performance.} 
Tab.~\ref{tab:performance} shows the runtime performance averaged for generating 100 samples. 
The measured metrics include: the total number of parameters (Param, M), the computational cost of one-time inference (GFLOPs), the inference time (s), the post-processing time averaged over successful samples (s), the memory consumption of intermediate variables in topology diffusion (T.Mem, MB), and the memory consumption of intermediate variables in geometry diffusion (G.Mem, MB).
\begin{table}[h]
    \centering
    \setlength{\tabcolsep}{3pt} 
    \caption{Runtime Performance. T.Mem refers to the memory used by topology, measured in MB. G.Mem refers to the memory used by geometry entities or latents. 
    }
    \label{tab:performance}
    \vspace{-3mm}
    \resizebox{\linewidth}{!}{
    \begin{tabular}{c|cccccc}
    \toprule
     & Params & GFLOPs & Inference & Postprocess & T.Mem & G.Mem \\ \midrule
    BrepGen                 &291.57 &83.69& 3.16 & \textbf{1.56} & N/A & 26.48\\ 
    DTGBrepGen              &145.21 &7.25 & 0.62 & 5.93 & 1.45 & 1.39 \\
    BrepDiff                & 259.71 & 3.51 & 0.34 & 15.56 & N/A & 0.088\\
    \hline
    Ours               &\textbf{82.18} & \textbf{2.20} & \textbf{0.21} & 2.34 & \textbf{0.058} & \textbf{0.043}\\
    \bottomrule
    \end{tabular}
    }
\end{table}

Our method significantly outperforms competing approaches in computational efficiency, requiring 1.76$\times$ fewer parameters, 1.59$\times$ fewer GFLOPs, and achieving a 1.85$\times$ reduction in total runtime.
BrepGen benefits from a relatively fast post-processing pipeline. The reason is that BrepGen’s post-processing relies on the network-generated results and follows a very simple process.
BrepDiff is also quite efficient in terms of inference time, however, it lags far behind others on post-processing time because of the reliance on surface reconstruction and mesh intersection. 
The intermediate memory consumption of our topology generation network (T.Mem) and geometry generation network (G.Mem) is the lowest among all methods. In addition, our method achieves a 615× improvement over BrepGen in geometry generation efficiency.
We further evaluate the reconstruction performance of the VAE. Compared to BrepGen, our method achieves a lower reconstruction error of 0.000003 (vs. 0.000011) on DeepCAD and 0.000013 (vs. 0.000016) on the ABC dataset.Although our VAE uses a latent space with three times the dimensionality of BrepGen, it achieves a final KL loss of 274.4 (vs. 74.2). After normalizing by the latent dimensionality (274.4/3 = 91.4), the KL loss remains higher than BrepGen (91.4 vs. 74.2), indicating a more informative latent representation.

\begin{figure}[h]
    \centering
    \includegraphics[width=1.0\linewidth]{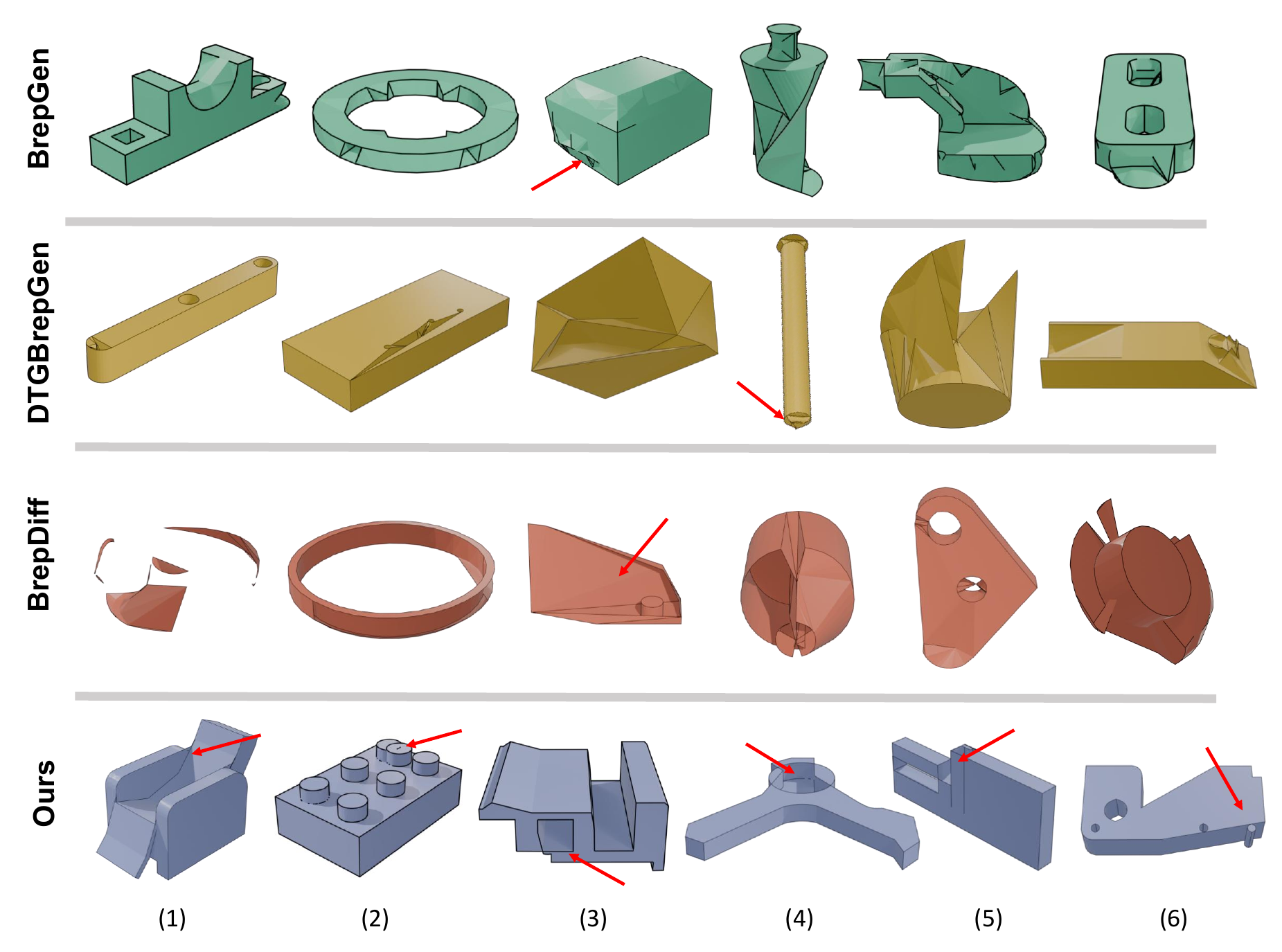}
    \caption{Comparison of failure cases.}
    \label{fig:exp_failures}
\end{figure}

\head{Failure Comparison.} 
Fig.~\ref{fig:exp_failures} illustrates failure cases of competing methods. BrepGen and DTGBrepGen jointly generate vertices, edges, and surfaces, making them prone to incorrect connectivity. 
Conversely, BrepDiff lacks explicit topological modeling, leading to missing surfaces and self-intersections. 
We manually evaluated 500 samples to assess how often cases that are considered valid by post-processing but actually invalid (Fig. 13) occur. The failure rates on DeepCAD-0-30 are BrepGen: 41.2\%, BrepDiff: 19.6\%, DTGBrepGen: 5.7\%, and ours: 3.1\%. 
Our approach can avoid the aforementioned failures to some extent, owing to the surface-centric representation and the topology-guided geometry generation. 
Overall, our method usually fails when generating complex shapes and may produce incorrectly trimmed surfaces and small redundant surfaces.

\subsection{Ablation Study}\label{subsec:ablate}
We conduct ablation studies on the DeepCAD-7-30 dataset. Results are shown in Tab.~\ref{tab:comparison}.
Replacing the generated topology with the ground truth topology (\emph{w/ GT topology}) achieves only a slightly better metric, indicating the high quality of the generated topology.
Including or excluding this component DiT-L/H does not significantly affect the results, validating that the method mainly contributes to efficiency improvements rather than performance gains. The lightweight design significantly reduces both the computational cost and the model size, decreasing the GFLOPs from 1.77 to 1.03 and the number of parameters from $87.1$M to $39.04$M.
We remove the GNN module from the VAE to evaluate the impact of surface-to-surface interactions within the VAE on the final results. The results show that removing this component leads to a slight performance degradation, with all metrics experiencing a minor decline.
BLISS is a preprocessing step for topology training, used to normalize topologies. We found that removing it significantly affects the final results. Under the same number of training epochs, the efficiency of topology learning drops substantially, leading to failures in generating complex topologies and causing the final outputs to be biased toward simpler topological structures.
We change the encoding scheme from treating an entire line of topological elements as a single token to encoding each individual element as a separate token. Under the same number of training iterations, all metrics decline, significantly impacting the final results. This demonstrates that our original design is indeed effective, as it increases information density and facilitates network learning.

\begin{table}[h]
    \centering
    \caption{Ablation experiment conducted on DeepCAD-7-30.}
    \label{tab:comparison}
    \vspace{-3mm}
    \setlength{\tabcolsep}{1mm}
    \begin{tabular}{c|cccc}
    \toprule
      & COV(\%) $\uparrow$ & MMD($10^{-2}$) $\downarrow$ & JSD($10^{-2}$) $\downarrow$ & Valid(\%) $\uparrow$ \\ \midrule
      Ours             & 76.4 & 1.22 & 0.87 & 83.6 \\ 
    + GT Topo          & 77.2 & 1.20 & 0.73 & 84.3 \\ 
    - DiT-H/L          & 76.3 & 1.21 & 0.85 & 83.9 \\
    - VAE-GNN          & 76.1 & 1.24 & 0.95 & 79.2 \\
    - BLISS            & 72.6 & 1.34 & 1.05 & 69.8 \\
    row -> element     & 73.7 & 1.30 & 1.01 & 72.4 \\
    \bottomrule
    \end{tabular}
\end{table}

As shown in Tab.~\ref{tab:pp_num}, on the DeepCAD-7-30 dataset, we analyze the impact of different numbers of post-processing retries on the success rate. The success rate achieved with three retries is already close to the highest success rate obtained with ten retries. We choose three retries to align with the three threshold retries used by the other baseline methods.
\begin{table}[h]
    \centering
    \caption{Comparison of the success rates under different numbers of post-processing retries.}
    \label{tab:pp_num}
    \vspace{-3mm}
    \resizebox{\linewidth}{!}{
    \setlength{\tabcolsep}{5mm}
    \begin{tabular}{c|cccc}
    \toprule
      & 1 & 2 & 3 & 10 \\ \midrule
      Valid(\%)             & 62.4 & 76.2 & 83.6 & 86.9 \\ 
    \bottomrule
    \end{tabular}
    }
\end{table}

\begin{figure}[h]
    \centering
    \includegraphics[width=\linewidth]{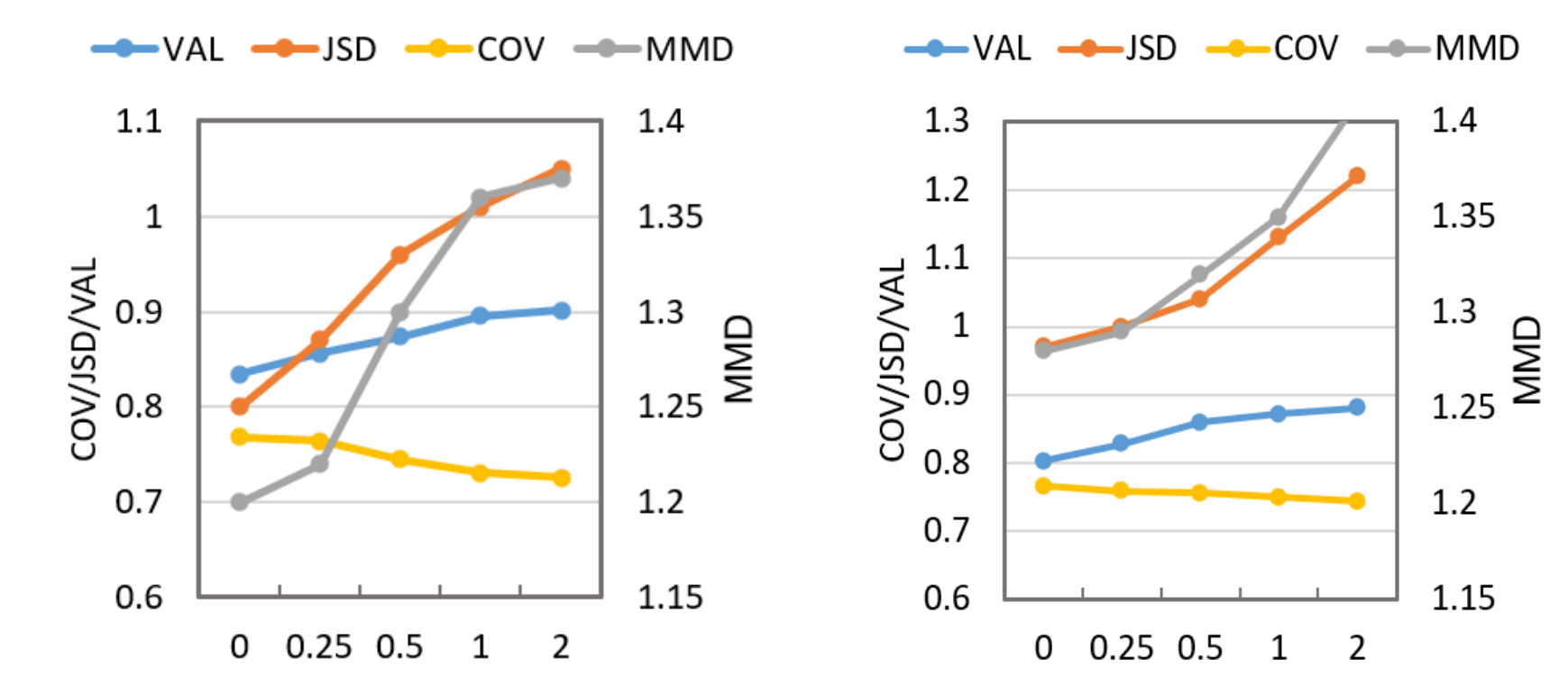}
    \caption{The impact of hyperparameters $m, n$ of Eq.(\ref{eq:scale}) on the performance. Left: varying $m$ with $n=0$; right: varying $n$ with $m=1$.}
    \label{fig:exp_hyp_v}
\end{figure}

We evaluate the impact of a scaling factor $\eta$ applied to the discrete log-probabilities during D3PM inference in Fig.~\ref{fig:exp_hyp_v}. Since D3PM outputs categorical distributions in log-space, $\eta$ functions as a temperature control: a higher $\eta$ sharpens the distribution to favor deterministic generation, while a lower $\eta$ flattens it to enhance structural diversity. 

Fig.~\ref{fig:exp_hyp_v} shows the impact of varying the topology sample parameters $m,n$ in Eq.(~\ref{eq:scale}). The overall trends of changing two parameters are similar but with a different slope. 
Increasing $m,n$ can significantly improve the success rate of geometric trimming, reaching a peak validity of 90.6\% on DeepCAD-7-30. 
However, excessively high confidence also leads to the deviation of the generated topologies from the true data distribution, leading to worse metrics of JSD, COV, and MMD. 
We select $m=0.8$ and $n=0.25$ as a balance between the generation validity and quality.

In addition, we compared the results of two-stage generation (our methods with both topology and geometry generation) and single-stage geometry generation on this dataset. The single-stage setting refers to geometry generation without topological guidance. The indicator results without post-processing are shown in Tab.~\ref {tab:single-vs-two}.
In terms of evaluation metrics, the two-stage generation consistently outperforms the single-stage setting across all metrics.

\begin{table}[h]
    \centering
    \caption{Comparison of single-stage and two-stage results on DeepCAD-7-30 without post-processing.}
    \label{tab:single-vs-two}
    \vspace{-3mm}
    \setlength{\tabcolsep}{1mm}
    \begin{tabular}{c|cccc}
    \toprule
      & COV*(\%) $\uparrow$ & MMD*($10^{-2}$) $\downarrow$ & JSD*($10^{-2}$) $\downarrow$ \\ \midrule
      single-stage          & 77.8 & 1.15 & 0.89  \\ 
      two-stage             & 79.5 & 1.01 & 0.78  \\ 
      
    \bottomrule
    \end{tabular}
\end{table}

\begin{table}[h]
    \centering
    \caption{Quantitative results without post-processing. (1) denotes generation without topology condition, while (2) denotes generation with topology condition.}
    \label{tab:one_two_1000}
    \setlength{\tabcolsep}{2mm}
    \begin{tabular}{c|cc|cc|cc}
    \toprule
    \multirow{2}{*}{Epochs} &
    \multicolumn{2}{c|}{COV*(\%) $\uparrow$} &
    \multicolumn{2}{c|}{MMD*($10^{-2}$) $\downarrow$} &
    \multicolumn{2}{c}{JSD*($10^{-2}$) $\downarrow$} \\
    & (1) & (2) & (1) & (2) & (1) & (2) \\
    \midrule
    100  & 72.1 & 74.2 & 1.55 & 1.24 & 2.81 & 2.41 \\
    200  & 73.4 & 75.4 & 1.42 & 1.22 & 2.03 & 1.82 \\
    500  & 75.0 & 76.9 & 1.30 & 1.18 & 1.72 & 1.37 \\
    1000 & 76.1 & 77.6 & 1.22 & 1.16 & 1.56 & 1.29 \\
    \bottomrule
    \end{tabular}
\end{table}
As shown in Tab.~\ref{tab:one_two_1000}, we compare the metrics at the early stage of training to demonstrate that topology conditioning accelerates convergence.



\begin{figure*}
    \centering
    \includegraphics[width=1.0\linewidth]{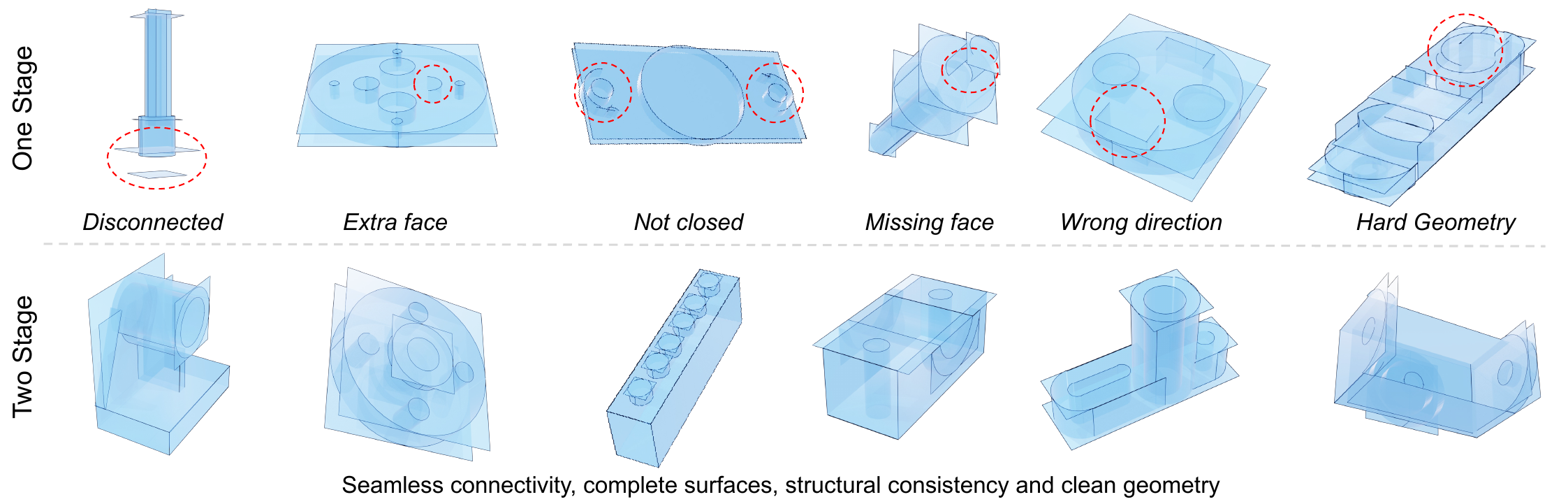}
    \caption{Comparison of generated patch results between the one-stage and two-stage settings.}
    \label{fig:wrong_compare}
\end{figure*}

Fig.~\ref{fig:wrong_compare} compares the results of single-stage and two-stage generation. Since diffusion models inherently favor distribution fitting rather than structural variation, the single-stage approach tends to produce CAD entities that resemble those in the training dataset. However, it struggles to accurately model each surface, often leading to issues such as disconnected components, redundant or missing surfaces, and structural inconsistencies. Moreover, its ability to generate complex geometries is significantly weaker than that of the two-stage method.
This motivates the introduction of topology as an explicit condition, which stabilizes training and improves the overall completeness of CAD generation. By incorporating topological guidance, the model can explicitly reason about inter-surface relationships rather than modeling the entire entity holistically.

\subsection{Generation Results on the Furniture Dataset}\label{subsec:other_gen}
We also conduct evaluations on the Furniture dataset. We compare against DTGBrepGen, the only prior method that reports results on this dataset. Due to the limited number of samples, we further perform a user study to demonstrate that our generated results are more complex.
The qualitative results are shown in Fig.\ref{fig:exp_furniture}.
\begin{table}[h]
    \centering
    \caption{Quantitative comparison of unconditioned B-rep generation on the furniture dataset.}
    \setlength{\tabcolsep}{4mm}
    \label{tab:furniture_result}
    \begin{tabular}{c|cc}
    \toprule
    {Furniture-0-50}  & Topo Valid$\uparrow$ & Valid(\%) $\uparrow$  \\ \midrule
    DTGBrepGen               & 92.1 & 64.2 \\
    \hline
    Ours                     & 95.8 & 71.2 \\
    \bottomrule
    \end{tabular}
\end{table}
We report the topological success rate and final geometric success rate on the Furniture dataset in Tab.~\ref{tab:furniture_result}.
Both methods use the original category distribution for sampling, and the final results are reported as averages. As shown in the table, our method achieves a higher success rate than DTGBrepGen.

\subsection{Topology–Geometry Inconsistency Metric}\label{subsec:TG-inconsistency}

As all successfully post-processed B-reps are constructed according to the predicted topology, they are topologically consistent by design. To evaluate the consistency before post-processing, we consider two generated surfaces as geometrically intersecting if they pass a proximity test with a predefined distance threshold, and measure whether the resulting intersections agree with the predicted topological adjacency. The resulting topology--geometry consistency is reported in Table~\ref{tab:topo_geo_consistency}. We further evaluate the generated topology by computing the Jensen--Shannon divergence (JSD) between the node degree distributions of the generated and ground-truth topologies (lower is better).

\begin{table}[h]
    \centering
    \caption{Topology-geometry consistency during geometry generation and topology distribution similarity.}
    \label{tab:topo_geo_consistency}
    \resizebox{\linewidth}{!}{
    \begin{tabular}{c|cc}
    \toprule
    Dataset &
    Topology--Geometry Consistency (\%) $\uparrow$ &
    Degree JSD $\downarrow$ \\
    \midrule
    DeepCAD-0-30 & 94.4 & 0.012 \\
    ABC-0-50     & 83.5 & 0.018 \\
    \bottomrule
    \end{tabular}
    }
\end{table}

\section{Limitations}\label{sec:limits}
Our method constructs watertight solids by extending surfaces to obtain intersection curves, and then computing intersection points from the intersections between these curves. Based on this, we design an algorithm to assemble a closed solid. However, this approach may introduce certain issues. When surfaces are extended too far, cases that should correspond to a single surface may instead produce multiple fragmented surfaces that are later stitched together. Although they are macroscopically equivalent, this can introduce additional internal edges, which are subsequently resolved through surface merging, at the cost of extra computational overhead.
In addition, due to limitations of the underlying OCC intersection kernel, intersection curves that are geometrically close and should theoretically intersect may still produce large distance values, causing the kernel to fail to generate valid intersection points. This numerical inconsistency effectively behaves like a geometric offset and therefore requires a relatively large tolerance to resolve the intersections.
Finally, the extension-based intersection strategy itself has inherent limitations. For geometries such as circular rings or spheres, even after extension, it can still be difficult to obtain meaningful intersection curves, which remains a major challenge of the method.

\section{Conclusions}\label{sec:concl}
We introduced \method{}, a two-stage approach that decouples complex B-rep generation into surface generation and surface adjacency modeling. 
By employing a two-stage diffusion process, our method alleviates the topological inconsistencies typically found in joint geometric modeling. 
Specifically, incorporating topology as a guiding condition enables the model to focus on global inter-surface relationships rather than isolated surface details. 
This leads to improved structural integrity and geometric complexity of the generation results.
Achieving state-of-the-art results on standard benchmarks (DeepCAD and ABC) with only 82.18M parameters, \method{} demonstrates a superior balance between generative capacity and architectural efficiency.
This work establishes a scalable paradigm for automated B-rep generation, bridging the gap between generative modeling and industrial designs.

\bibliographystyle{ACM-Reference-Format}
\bibliography{brep}





\end{document}